\documentclass[useAMS,usenatbib]{mn2e}

\usepackage{graphicx}
\usepackage{amssymb}
\usepackage{multirow}
\usepackage{pifont}
\usepackage{amsmath}

\makeatletter
    
    \newcommand{\Rmnum}[1]{\expandafter\@slowromancap\romannumeral #1@}
\makeatother

\title[Apsidal asymmetric-alignment of Jupiter Trojans]
         {Apsidal asymmetric-alignment of Jupiter Trojans}
\author[Jian Li, Hanlun Lei and  Zhihong J. Xia]
{Jian Li$^{1,2}$\thanks{E-mail: ljian@nju.edu.cn}, Hanlun Lei$^{1,2}$ and  Zhihong J. Xia$^{3}$\thanks{E-mail: xia@math.northwestern.edu}\\
$^1$School of Astronomy and Space Science, Nanjing University, 163 Xianlin Avenue, Nanjing 210023, PR China,\\
$^2$ Key Laboratory of Modern Astronomy and Astrophysics in Ministry of Education, Nanjing University, Nanjing 210023, PR China\\
$^3$Department of Mathematics, Northwestern University, 2033 Sheridan Road, Evanston, IL  60208, USA}
\begin{document}

\date{Accepted 1988 December 15. Received 1988 December 14; in original form 1988 October 11}

\pagerange{\pageref{firstpage}--\pageref{lastpage}} \pubyear{2002}

\maketitle

\label{firstpage}

\begin{abstract}

The most distant Kuiper belt objects exhibit the clustering in their orbits, and this anomalous architecture could be caused by Planet 9 with large eccentricity and high inclination. We then suppose that the orbital clustering of minor planets may be observed somewhere else in the solar system. In this paper, we consider the over 7000 Jupiter Trojans from the Minor Planet Center, and find that they are clustered in the longitude of perihelion $\varpi$, around the locations  $\varpi_{\mbox{\scriptsize{J}}}+60^{\circ}$ and $\varpi_{\mbox{\scriptsize{J}}}-60^{\circ}$ ($\varpi_{\mbox{\scriptsize{J}}}$ is the longitude of perihelion of Jupiter) for the L4 and L5 swarms, respectively. Then we build a Hamiltonian system to describe the associated dynamical aspects for the co-orbital motion. The phase space displays the existence of the apsidally aligned islands of libration centered on $\Delta\varpi=\varpi-\varpi_{\mbox{\scriptsize{J}}}\approx\pm60^{\circ}$, for the Trojan-like orbits with eccentricities $e<0.1$. Through a detailed analysis, we have shown that the observed Jupiter Trojans with proper eccentricities $e_p<0.1$ spend most of their time in the range of $|\Delta\varpi|=0-120^{\circ}$, while the more eccentric ones with $e_p>0.1$ are too few to affect the orbital clustering within this $\Delta\varpi$ range for the entire Trojan population. Our numerical results further prove that, even starting from a uniform $\Delta\varpi$ distribution, the apsidal alignment of simulated Trojans similar to the observation can appear on the order of the age of the solar system. We conclude that the apsidal asymmetric-alignment of Jupiter Trojans is robust, and this new finding can be helpful to design the survey strategy in the future.

 \end{abstract}


%_____________________________________________________________________________________________________________________
\section{Introduction}

Presently there are 14 extreme Kuiper belt objects (KBOs) discovered with semimajor axes exceeding 250 AU \citep[e.g.,][]{brow04, chen13, truj14, shep16}. It was noted that all these objects are clustered both in the longitude of perihelion and in the longitude of ascending node, and the confidence level can be as high as $99.8\%$ \citep{brow19}. This distinct orbital feature leads to the existence of an additional planet, referred to as Planet 9, which resides on a highly eccentric and inclined orbit \citep{baty16, baty19}. The secular perturbation of Planet 9 can be a dynamical mechanism which forces the orbital alignment of the known extreme KBOs, as well as the effect of its mean motion resonances (MMRs) \citep{beus16}.

We then speculate that the similar orbital alignment could possibly be observed elsewhere much closer to the Sun in the solar system. For instance, for some known planet moving on the eccentric orbit, it could also confine the apsidal configuration of the objects trapped in its MMRs. The first planet came into our thought is Jupiter, which has the largest mass and a relatively large eccentricity ($e_{\mbox{\scriptsize{J}}}=0.05$) among the four outer planets. Thus we intend to reinvestigate the orbital structure of objects in the low order MMRs with Jupiter.

The Jupiter Trojans share the semimajor axis of Jupiter, around the leading (L4) or trailing (L5) triangular Lagrangian point, and they are said to be settled in the 1:1 mean motion resonance (MMR) with Jupiter. This population comprising thousands of observed asteroids is currently the second largest group of minor planets in the solar system, only fewer than the main belt asteroids. As such a large group of small bodies, they may serve as a good field experiment for the dynamical theories developed for Planet 9. It is also well known that the 2:1 Jovian MMR contains a population of asteroids, i.e., the so-called Hecuba gap centered at $\approx3.27$ AU \citep{schw69}. Although the number of this resonant population has increased to several hundreds, merely 124 objects (i.e., Zhongguos) were identified on stable orbits by \citet{chre15}. So these stable Zhongguos are too few to be used for the statistical analysis at present, and their orbital distribution would not be discussed in this paper.

Over the past decade, many studies have been done for better understanding of the physical and orbital distributions of the Jupiter Trojans. The Wide-field Infrared Survey Explorer (WISE) project has provided the most complete measures of physical properties for approximately 1900 Jupiter Trojans down to $\sim10$ km \citep{grav11, grav12}. According to the size and color observations, the Trojans are proposed to originate from the same primordial population as the Kuiper belt objects \citep{fras14, wong16}. While as for the albedo distribution, this property shows no statistical difference between the L4 and L5 swarms, probably indicating the identical chemical and dynamical evolutions \citep{emer11}. A more exhaustive review can be found in \citet{emer15}.

The known Jupiter Trojans have been better characterized in terms of their orbital features, and accordingly they were thought to be captured into the Lagrangian regions from the primordial planetesimal disk during the mass growth of Jupiter, by the dissipative forces such as gas drag or collisions \citep{shoe89, kary95, marz98, flem00, kort01}. However, this process can not explain the broad inclination distribution of the Jupiter Trojans up to $40^{\circ}$, which was later reproduced in the framework of the Nice model \citep{tsig05}. Once Jupiter and Saturn crossed their mutual 1:2 MMR, there was a period of time that the planetesimals can be scattered and trapped around L4 and L5 through chaotic paths, and acquire large inclinations; as well, the observed orbital distributions of the eccentricity and libration amplitude can be successfully generated \citep{morb05}.

Nevertheless, an outstanding issue remains, i.e., the number difference between the two Trojan swarms. With the samples detected by the WISE, \citet{grav11} estimated a number ratio of $N(\mbox{L}4)/N(\mbox{L}5)\sim1.4\pm0.2$. As a matter of fact, even the possible selection effects are involved, there are still significantly more Trojans in the L4 swarm \citep{szab07}. Recently, \citet{disi19} calculated the difference in the escape rate between the L4  and L5 Trojans, which is only as small as $\sim2\%$ over the age of the solar system. In additional, \citet{hell19} considered the Yarkovsky force on the Jupiter Trojans, since only objects with radii $\leqslant1$ km are significantly influenced, the number asymmetry existing at large size can not be explained. Consequently, the source of the asymmetry problem should be due to the initial capture implantation, but not the later long-term dynamical evolution. For instance, the Jumping Jupiter and capture mechanism started at $\sim3-10$ Myr after the birth of the Sun, could potentially create a number ratio of $N(\mbox{L}4)/N(\mbox{L}5)=1.3-1.8$ \citep{nesv13}.

Since the small Trojans with absolute magnitudes $H>11$ are far from observationally complete, their intrinsic distributions need to be further improved by increasing the number of the observed samples. According to our previous work \citep{li2018}, the total number of Jupiter Trojans with diameters $>1$ km (i.e., $H\lesssim19$) is estimated to be approximately half a million. This suggests that there is a huge Trojan population yet to be discovered in future surveys. Hence it is of special interest to investigate that whether the orbital orientation of the Jupiter Trojans has some particular characteristic. As we know, in the papers published so far for the various distributions of the Jupiter Trojans, the longitude of perihelion and ascending node have not been explicitly considered. If the clustering of either of these two orbital orientations can be detected, it can directly place constrains on preparing the future observation plans.

The rest of this paper is organized as follows. In Section 2,  we reveal the apsidal asymmetric-alignment of the observed Jupiter Trojans. In Section 3, we develop the theoretical approach to understand the dynamical mechanism that accounts for this anomalous orbital structure of the Jupiter Trojans, and correspondingly a detailed explanation is provided. In Section 4, we perform numerical simulations to reproduce the apsidally aligned Trojans, sculpted by the eccentric orbit of Jupiter, from an initially uniform distribution of the longitudes of perihelia. The conclusions and discussion are given in Section 5.

%_____________________________________________________________________________________________________________________

\section[]{Orbital orientations of observed Jupiter Trojans}

\begin{figure*}
  \centering
  \begin{minipage}[c]{1\textwidth}
  \vspace{0 cm}
  \includegraphics[width=9cm]{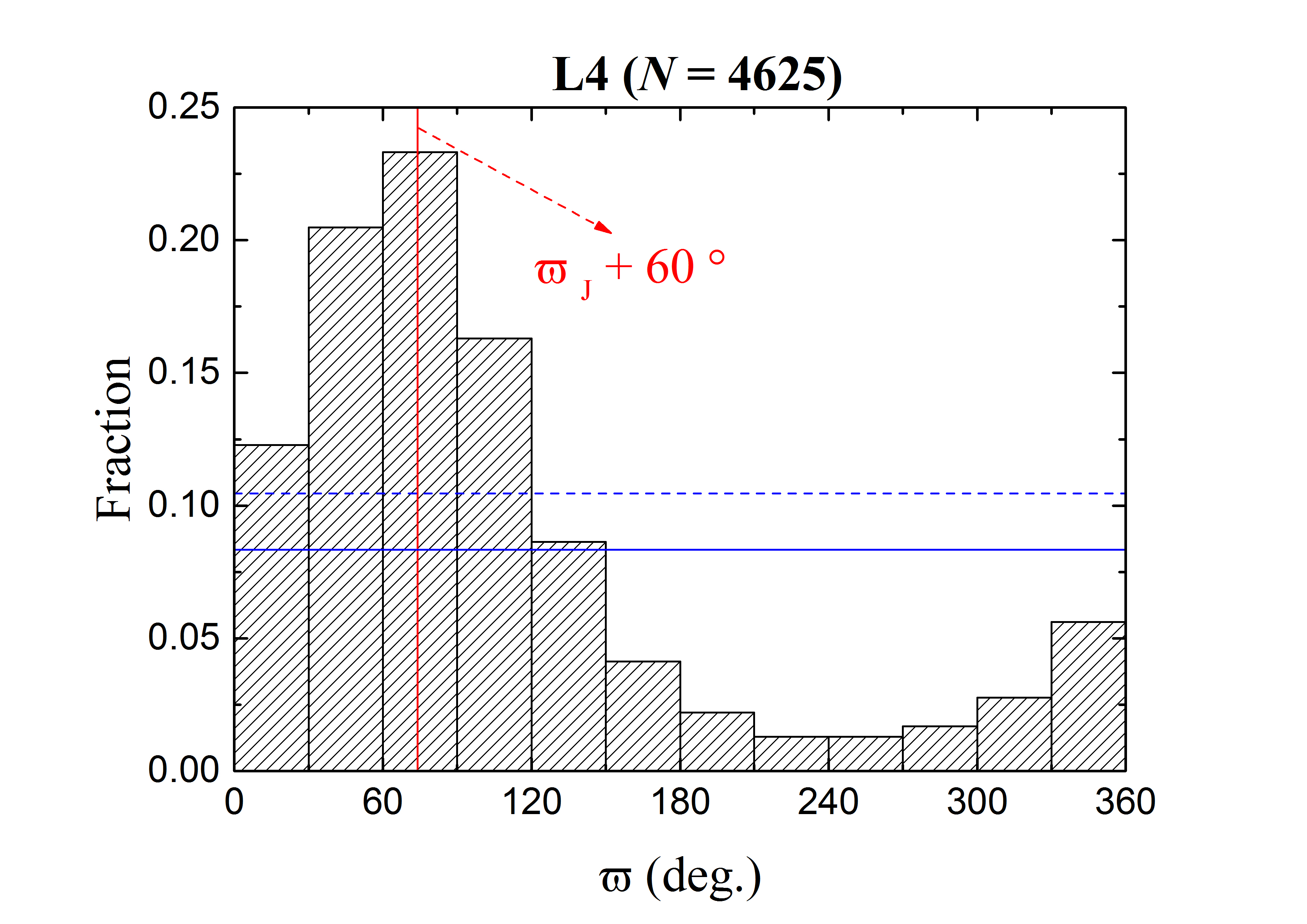}
  \includegraphics[width=9cm]{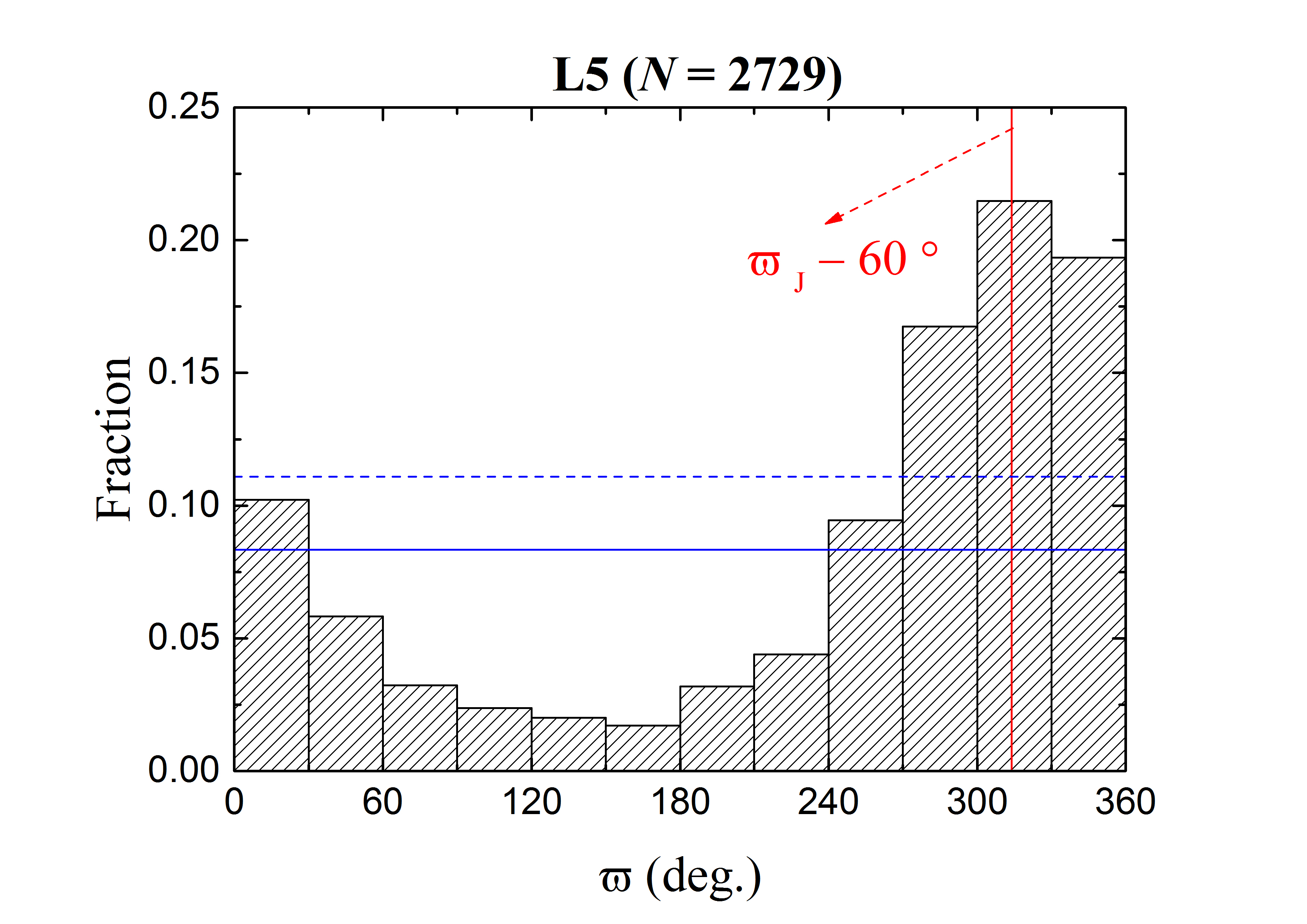}
  \end{minipage}
  \begin{minipage}[c]{1\textwidth}
  \vspace{0 cm}
  \includegraphics[width=9cm]{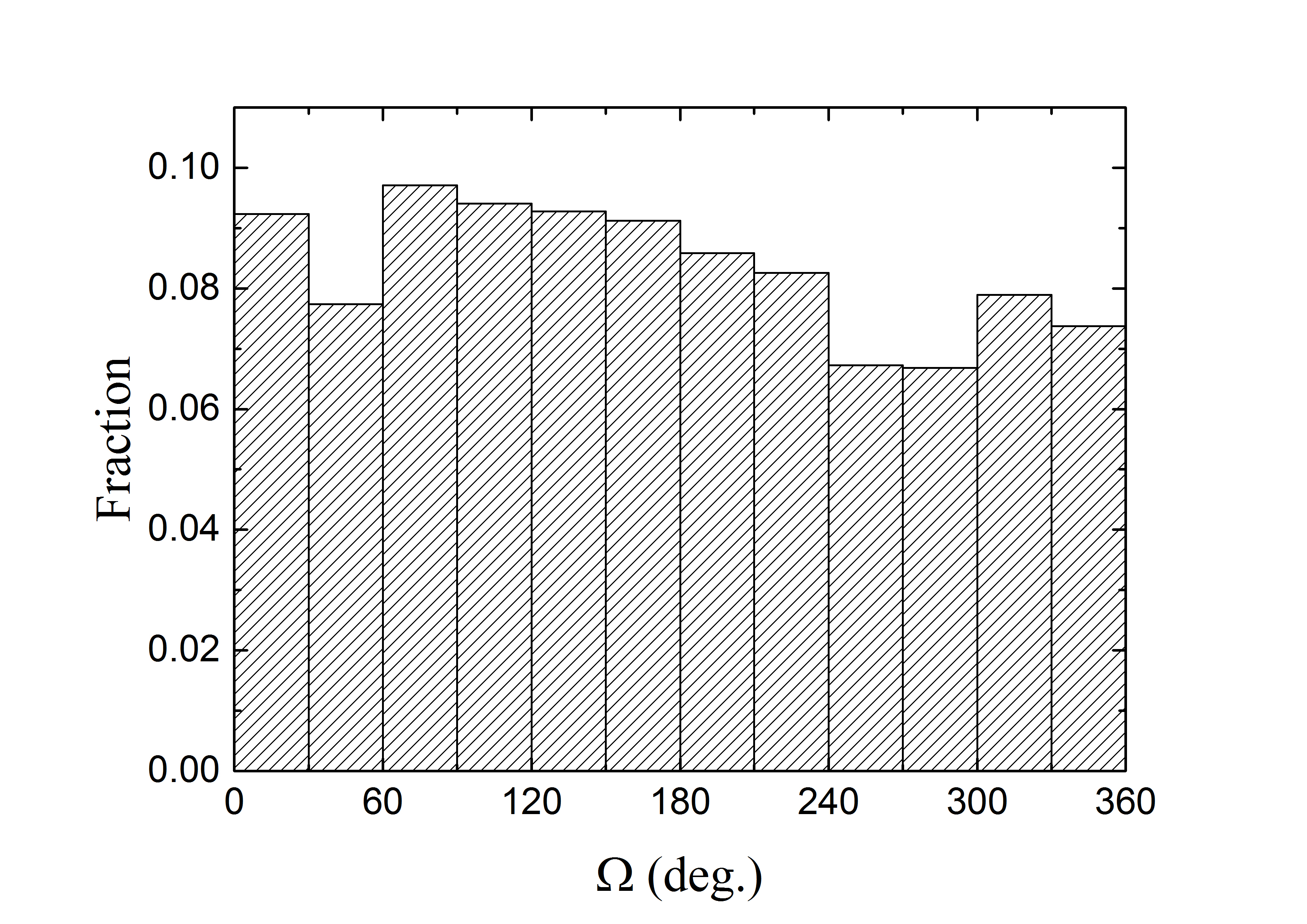}
  \includegraphics[width=9cm]{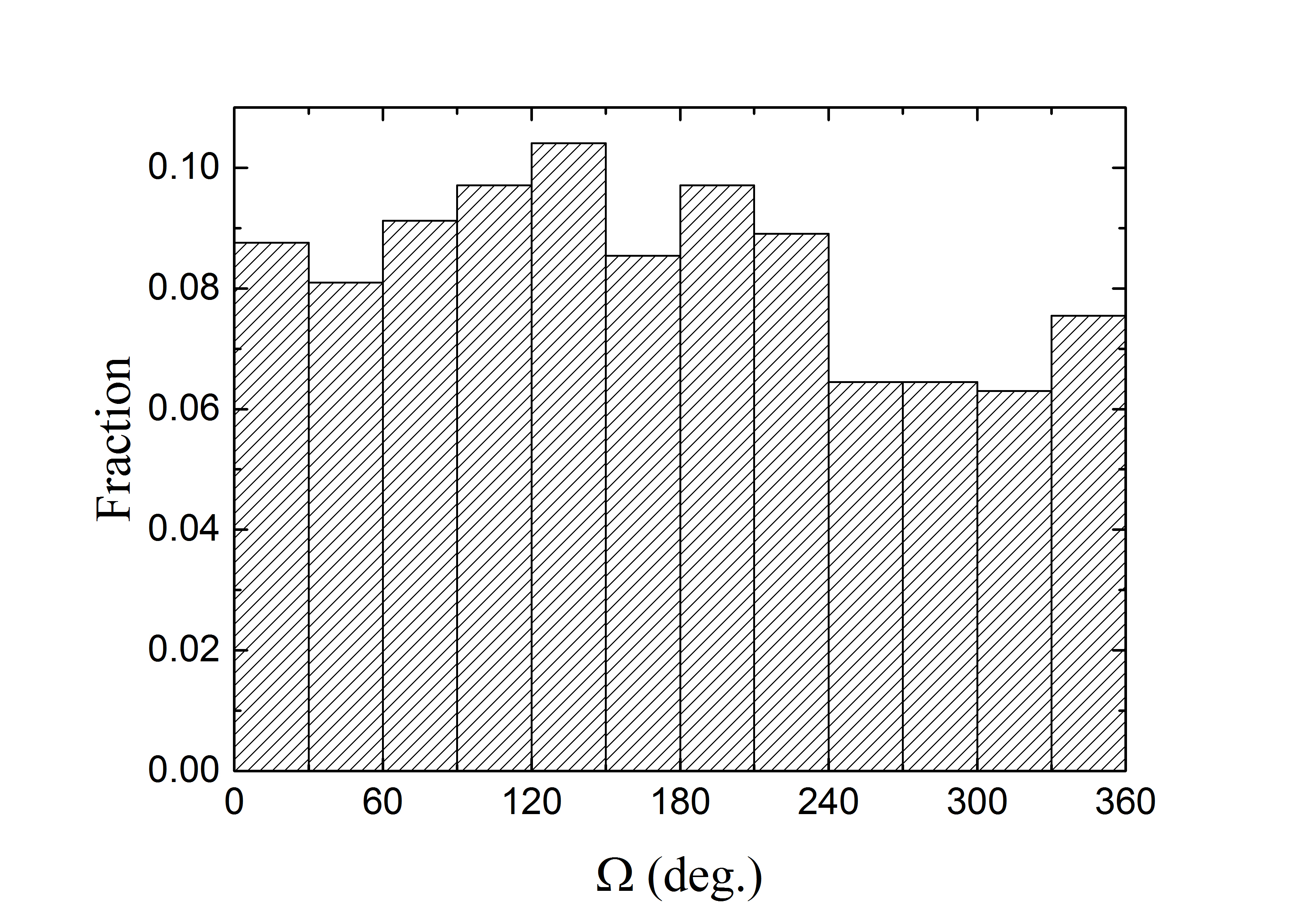}
  \end{minipage}
   \vspace{0 cm}
  \caption{Distribution of the longitudes of perihelia $\varpi$ (upper panels) and the longitudes of ascending nodes $\Omega$ (lower panels) of the observed Jupiter Trojans, at the epoch of 2020 May 31. (Left column) There are 4625 objects in the L4 region, and the values of their $\varpi$ are clustered around $\varpi_{\mbox{\scriptsize{J}}}+60^{\circ}$ ($\varpi_{\mbox{\scriptsize{J}}}$ is the longitude of perihelion of Jupiter). (Right column) There are 2729 objects in the L5 region, and the  $\varpi$ clustering is around $\varpi_{\mbox{\scriptsize{J}}}-60^{\circ}$. For reference, the horizontal solid lines are plotted in the upper two panels for the uniform distribution in $\varpi$, and the dashed lines indicate the 5$\sigma$ upper uncertainty from Poisson statistics.}
 \label{Obs}
 \end{figure*}

 \begin{figure*}
  \centering
  \begin{minipage}[c]{1\textwidth}
  \vspace{0 cm}
  \includegraphics[width=9cm]{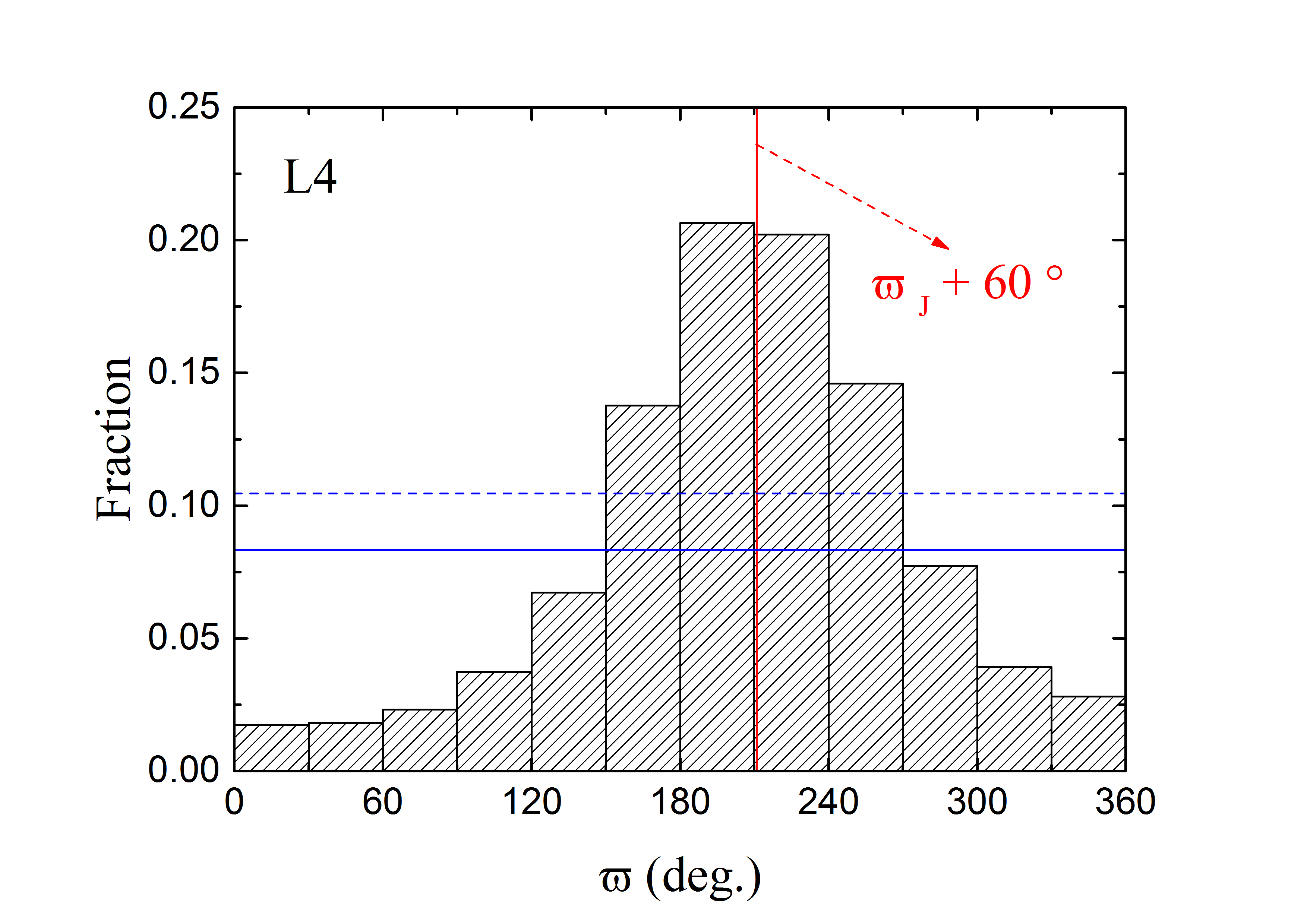}
  \includegraphics[width=9cm]{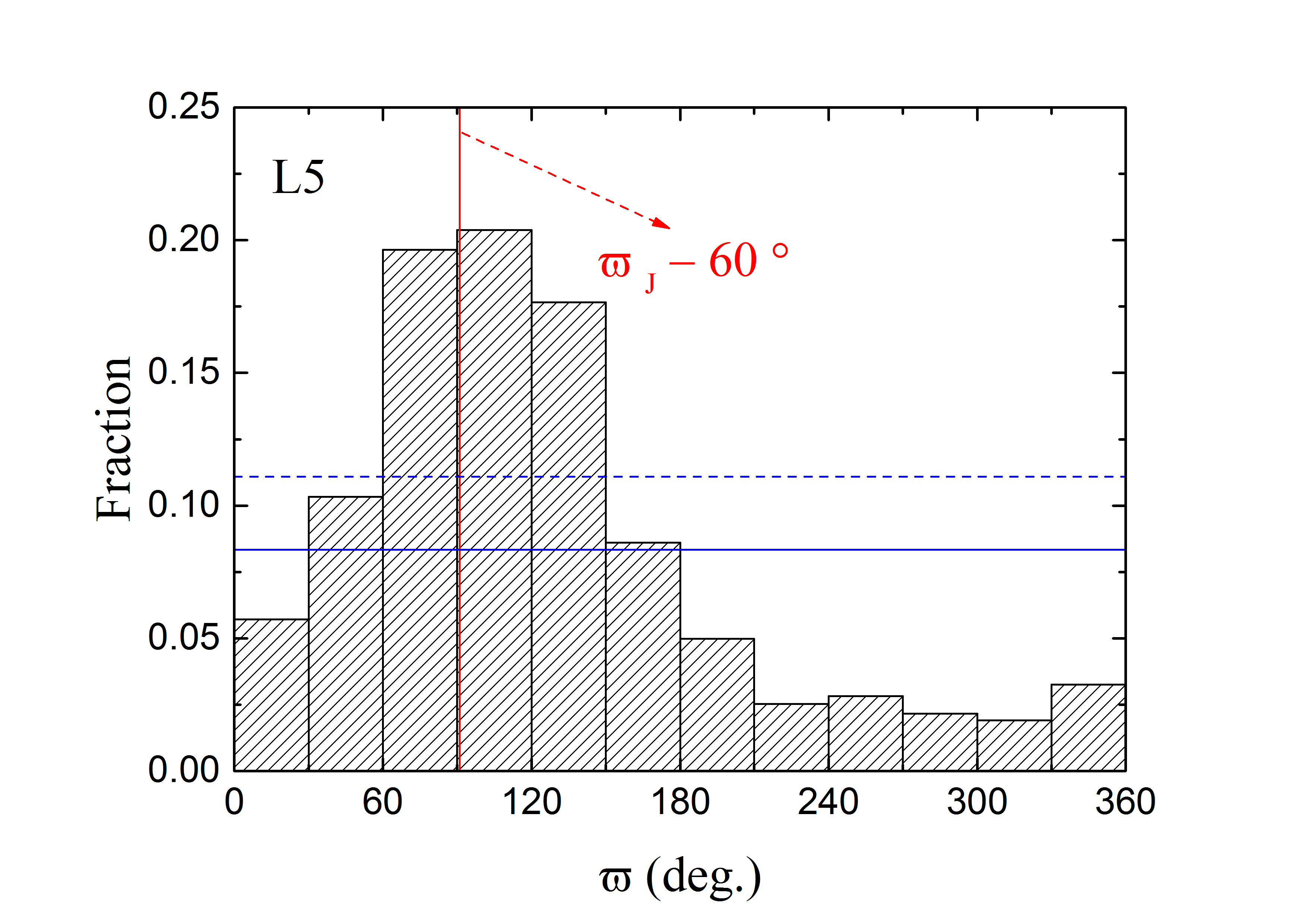}
  \end{minipage}
   \vspace{0 cm}
  \caption{As in the upper two panels of Fig. \ref{Obs}, but for the orbits of the observed Jupiter Trojans after 1 Myr evolution.}
 \label{ObsMyr}
 \end{figure*}  
 
As of 2020 May 27, there are more than 8100 Jupiter Trojans have been registered in the Minor Planet Center\footnote{https://minorplanetcenter.net/iau/lists/JupiterTrojans.html} (MPC). This population includes many objects that only have short-arc observations which may induce inaccurate orbit determinations for them. Thus we only select the Trojans that have been observed over multiple oppositions, resulting 7354 in total. To represent the true members of the Jupiter Trojans, we then numerically evolve the orbits of these samples under the perturbations of four planets in the outer solar system. The planets' masses, initial positions, and velocities are adopted from DE405, in the heliocentric frame referred to the J2000.0 ecliptic plane at epoch 1969 June 28 \citep{stan98}. Then the planets are integrated to the specified epoch of the Trojans. Finally, we compute the orbital evolution of the considered Trojans\footnote{Here and later in Sections. 3.2 and 4, the SWIFT\_RMVS3 symplectic integrator \citep{levi94} is used to numerically compute the orbital evolution of the Trojans. We adopt a time step of 0.5 year, which is about 1/24 of the shortest orbital period (i.e., Jupiter's period) in our model.}. At the end of the 1 Myr integration, all of them can survive around the individual Lagrangian points. Then these 7354 multioppositional and stable Jupiter Trojans could be used for our analysis. In this way, we can detect and exclude the possible transient Jupiter Trojans such as P/2019 LD2 which only can stay on Trojan-like orbit for about ten years \citep{hsie21}.

Fig. \ref{Obs} shows the orbital element distributions of the L4 (left column, 4625 in total) and L5 (right column, 2729 in total) populations. It can be seen in the upper two panels that, for either the L4 or L5 swarm of Jupiter Trojans, the clustering in the longitude of perihelion $\varpi$ is readily apparent. A simple way to quantify this apsidal alignment is to compare with the number fraction $f_{even}$ of an even $\varpi$ distribution. If we have $N$ Trojans in total, since the width of each $\varpi$ bin is adopted to be $30^{\circ}$, a sample size of $N_{bin}=(30/360)N$ per bin is expected, i.e., the mean fraction $f_{even}=N_{bin}/N\approx0.083\%$. Considering the random variations about the value of $f_{even}$ due to counting statistics, the uncertainty from Poisson distribution can be calculated by $\sigma=\pm f_{even}\cdot\sqrt{N_{bin}}/N_{bin}$. Accordingly, in the upper panels of Fig. \ref{Obs}, the horizontal solid and dashed lines are plotted to indicate the mean fraction $f_{even}$ and its 5$\sigma$ upper uncertainty, respectively. As the peak of the histograms can achieve the value of $\sim0.21$~(L5)~$-$~0.23~(L4), much higher than the dashed line, the clustering of  $\varpi$ shows strong significance over $5\sigma$ confidence level. Another key point shown in these two figures is that, the number peak corresponds to the bin of  $\varpi=30^{\circ}-90^{\circ}$ ($300^{\circ}-330^{\circ}$) for the L4 (L5) Trojan swarm, giving a difference of around $+60^{\circ}$ ($-60^{\circ}$) with Jupiter's longitude of perihelion $\varpi_{\mbox{\scriptsize{J}}}$ (indicated by the red line). 

%\textbf{If we have $N$ Trojans in total, since the width of each $\varpi$ bin is adopted to be $30^{\circ}$, a sample size of $N_{bin}=(30/360)N$ per bin is expected, i.e., the mean fraction $f_{even}=N_{bin}/N\approx0.083\%$. Considering the random variations about the value of $f_{even}$ due to counting statistics, the uncertainty from Poisson distribution can be calculated by $\sigma=\pm f_{even}\cdot\sqrt{N_{bin}}/N_{bin}$. Accordingly, in the upper panels of Fig. \ref{Obs}, the horizontal solid and dashed lines are plotted to indicate the mean fraction $f_{even}$ and its 2$\sigma$ uncertainty, respectively. As the peak of the histograms can achieve the value of $\sim0.21$~(L5)~$-$~0.23~(L4), much higher than the upper dashed line, the clustering of  $\varpi$ shows strong significance at $2\sigma$ confidence level, and could be actually over $10\sigma$ level.}

%Considering there are $N$ Trojans in total, since the width of each $\varpi$ bin is adopted to be $30^{\circ}$, a sample size of $N_{bin}=(30/360)N$ per bin is expected. Then the mean fraction $f_{even}$ can be calculated as $N_{bin}/N\approx0.083\%$, and the 1-$\sigma$ uncertainty from Poisson statistics is $\pm f_{even}\cdot1/\sqrt{N_{bin}}$. 

In order to further confirm the newly found characteristic of apsidal clustering, we need to consider the evolving orbits of Jupiter Trojans at different epoch. Since the periodicity in the variation of Jupiter's longitude of perihelion $\varpi_{\mbox{\scriptsize{J}}}$ is $\sim0.17$ Myr, we propose that the 1 Myr  timescale is long enough to check the persistence of the apsidal confinement of Jupiter Trojans due to different precession rates of their $\varpi$. Then the $\varpi$ distribution of the Jupiter Trojans after 1 Myr evolution is shown in Fig. \ref{ObsMyr}, from the integration mentioned at the beginning of this section. One can immediately notice the similar patterns depicted in Fig. \ref{Obs}: (1)  the longitudes of perihelia $\varpi$ are clustered around $\varpi_{\mbox{\scriptsize{J}}}+60^{\circ}$ and $\varpi_{\mbox{\scriptsize{J}}}-60^{\circ}$ for the L4 and L5 swarms, respectively; (2) the number fraction of the clustered samples within a $\varpi$ bin could be over 0.2, which is significantly larger than $f_{even}\sim0.083$ ($>5\sigma$ confidence level) in the case of a uniform $\varpi$ distribution. Therefore, the apsidal asymmetric-alignment of Jupiter Trojans should be intrinsic but not due to the observational bias. We suppose that the gravitational perturbation of Jupiter with eccentricity $e_{\mbox{\scriptsize{J}}}\sim0.05$ may induce and maintain this intriguing orbital distribution, and an explanation will be offered in the next section.

However, the longitudes of ascending nodes ($\Omega$) of Jupiter Trojans are not visually clustered in any $\Omega$ bin, as shown in the lower panels of Fig. \ref{Obs}. The element $\Omega$ measures the azimuthal direction of an object's orbital plane. Since Jupiter has a very low inclination, which is only about $0.3^{\circ}$ relative to the invariable plane of the solar system, this planet could not induce a significant clustering of the orbital planes of its Trojan population. Then it is quite understandable that there appears such a nearly uniform $\Omega$ distribution of Jupiter Trojans.

%_____________________________________________________________________________________________________________________

\section{Apsidal clustering: analysis}

\subsection{Theoretical approach}

In this subsection, to explore the existence of the apsidally aligned libration islands in the phase space, we consider the spatial elliptical restricted three-body problem (SERTBP) that consists of the Sun, Jupiter and a Jupiter Trojan. Since Jupiter Trojans reside in the 1:1 MMR with Jupiter, for such a co-orbital motion, the classical theoretical approach via the expansion of the disturbing function would fail \citep[see][chap.~6]{murr99}. Nevertheless, some new analytical treatments have been developed, such as the secular theories in \citet{mora99} and \citet{namo99}, the perturbative scheme in \citet{nesv02}, and the symplectic mapping model in \citet{lhot08}. Here we adopt a semi-analytic way to evaluate the resonant disturbing function with no expansion. We refer the reader to \citet{gall06} for more details on the following calculations.

\begin{figure*}
  \centering
  \begin{minipage}[c]{1\textwidth}
  \vspace{0 cm}
  \includegraphics[width=9cm]{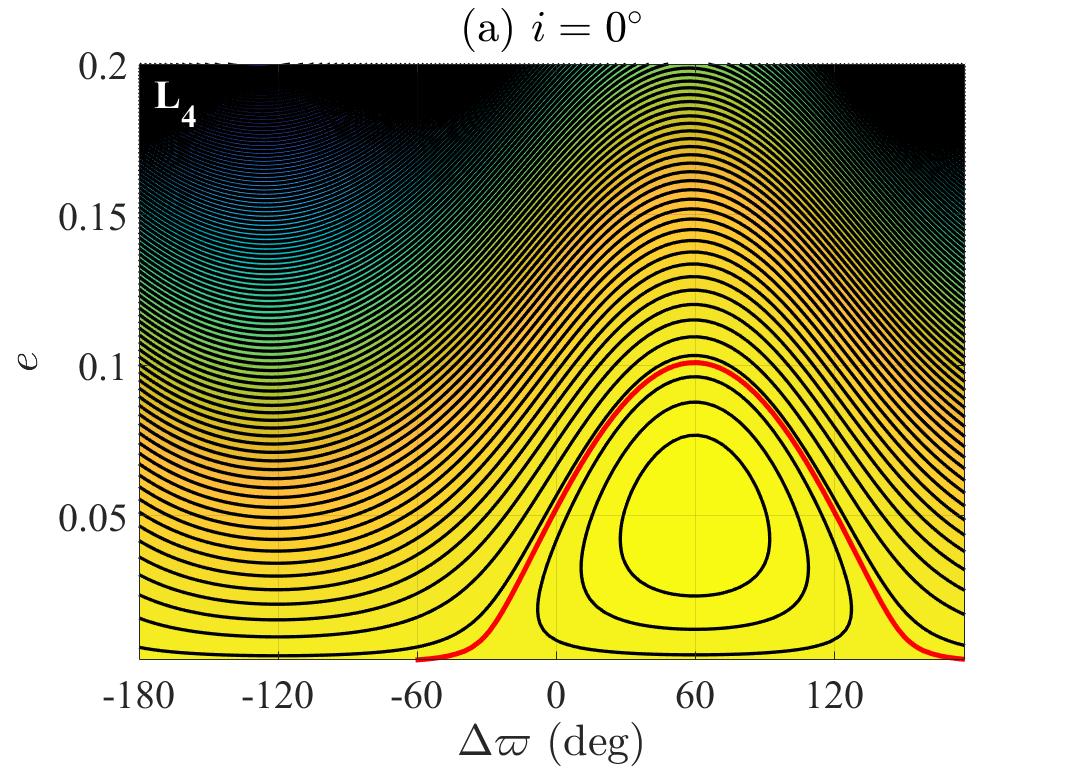}
  \end{minipage}
     \vspace{0 cm}
 \begin{minipage}[c]{1\textwidth}
  \vspace{0.2 cm}
  \includegraphics[width=9cm]{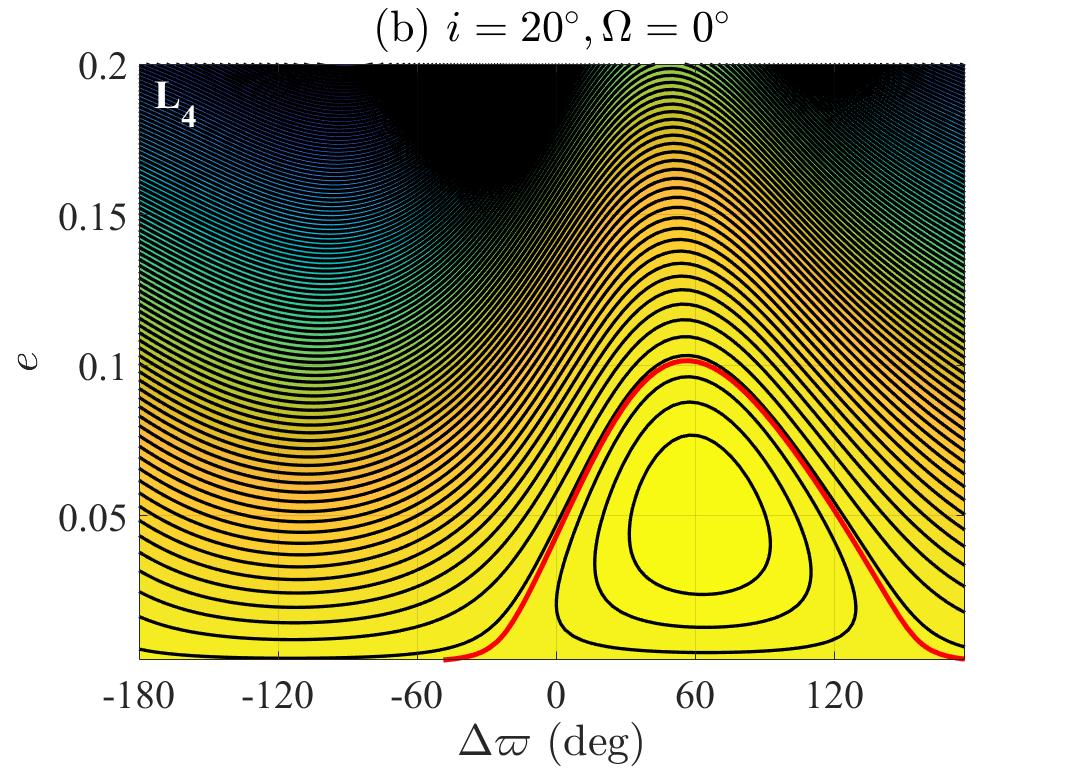}
  \includegraphics[width=9cm]{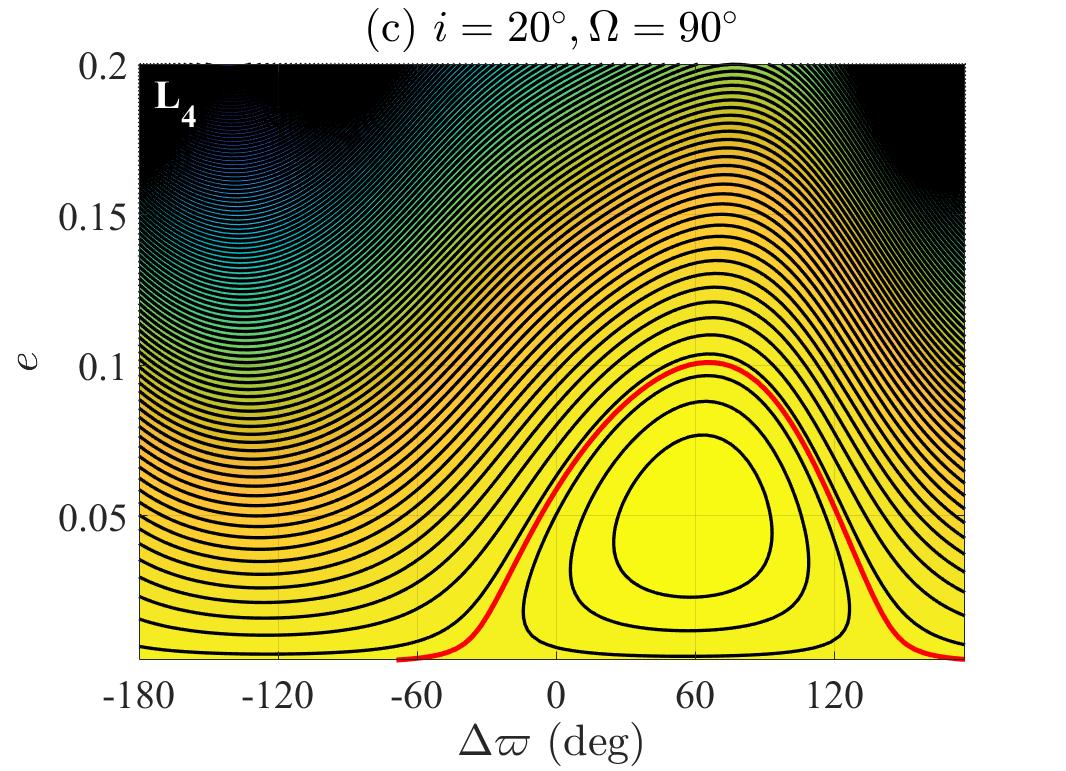}
  \end{minipage}
     \vspace{0 cm}
\begin{minipage}[c]{1\textwidth}
  \vspace{0.2 cm}
  \includegraphics[width=9cm]{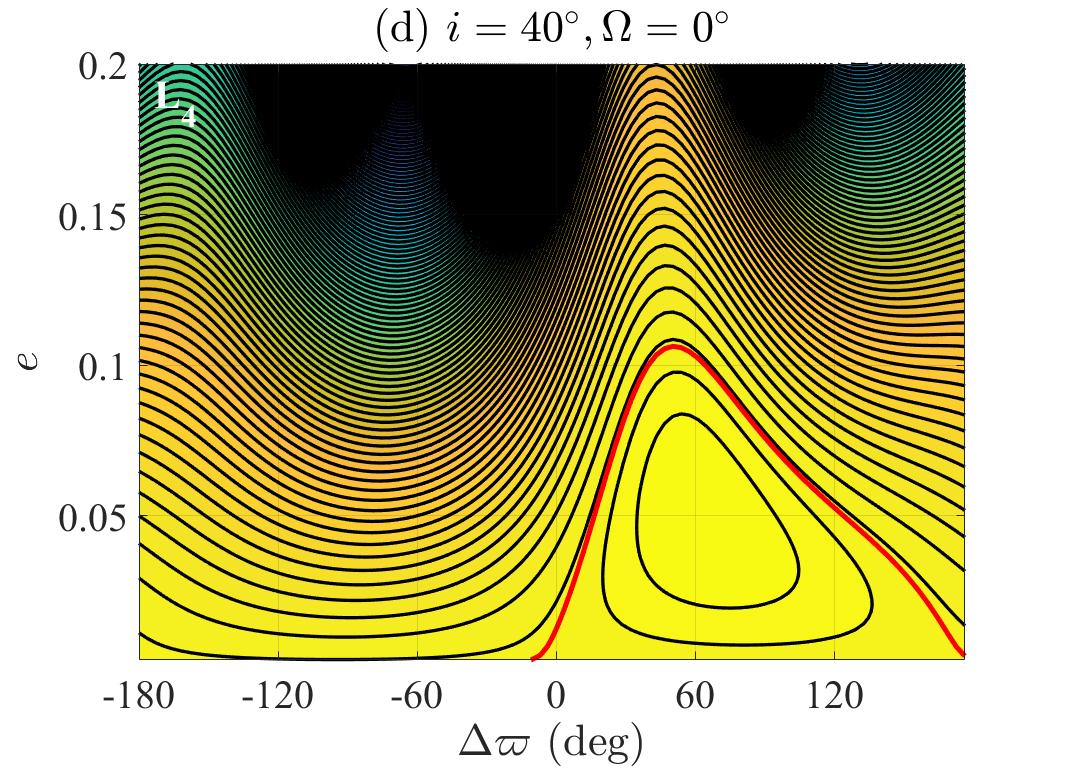}
  \includegraphics[width=9cm]{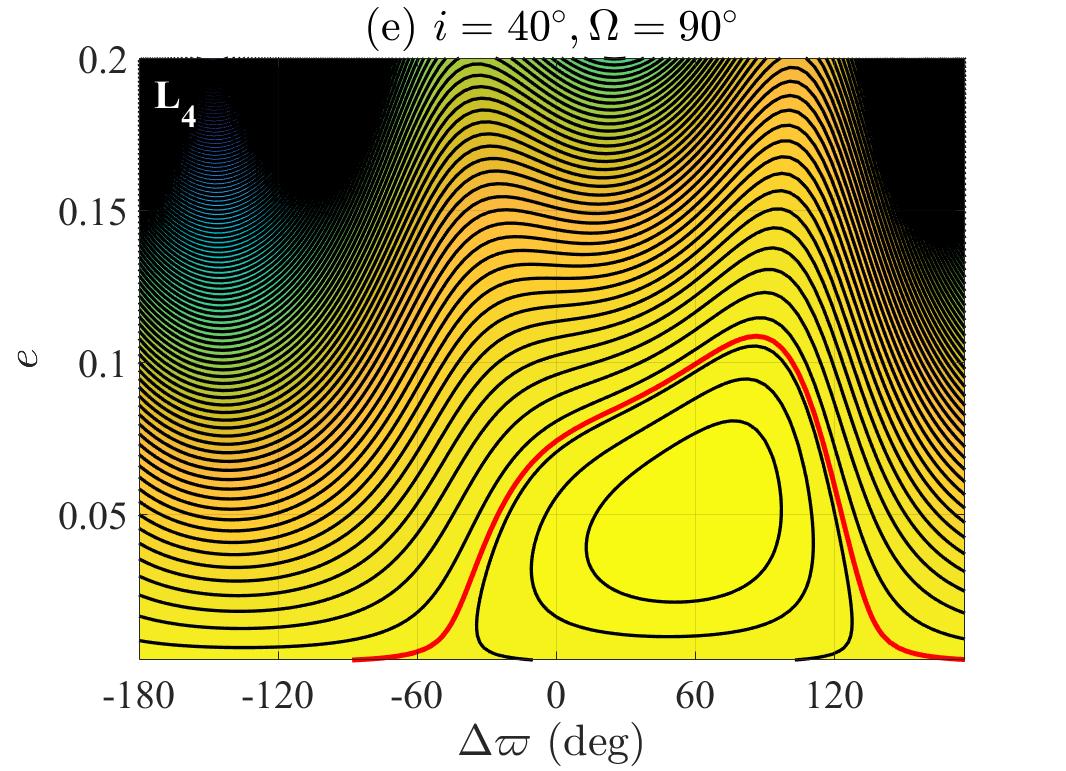}
  \end{minipage}
     \vspace{0 cm}
  \caption{The phase space structure near Jupiter's 1:1 MMR for the Lagrangian point L4, for the planar case of $i=0$, and the inclined cases of $i=20^{\circ}, 40^{\circ}$ and $\Omega=0, 90^{\circ}$ (see the labels in individual panels). It shows that there is always a stable equilibrium point around $\Delta\varpi=\varpi-\varpi_{\mbox{\scriptsize{J}}}=60^{\circ}$ for the L4 Trojan swarm. The red curve indicates the separatrix dividing the libration and circulation regions. In the calculation of the Hamiltonian by equation (\ref{Eq7}), the eccentricity $e_{\mbox{\scriptsize{J}}}$ of Jupiter is taken to be its real value of 0.05.}
 \label{theory1}
 \end{figure*}

 \begin{figure*}
  \centering
  \begin{minipage}[c]{1\textwidth}
  \vspace{0 cm}
  \includegraphics[width=9cm]{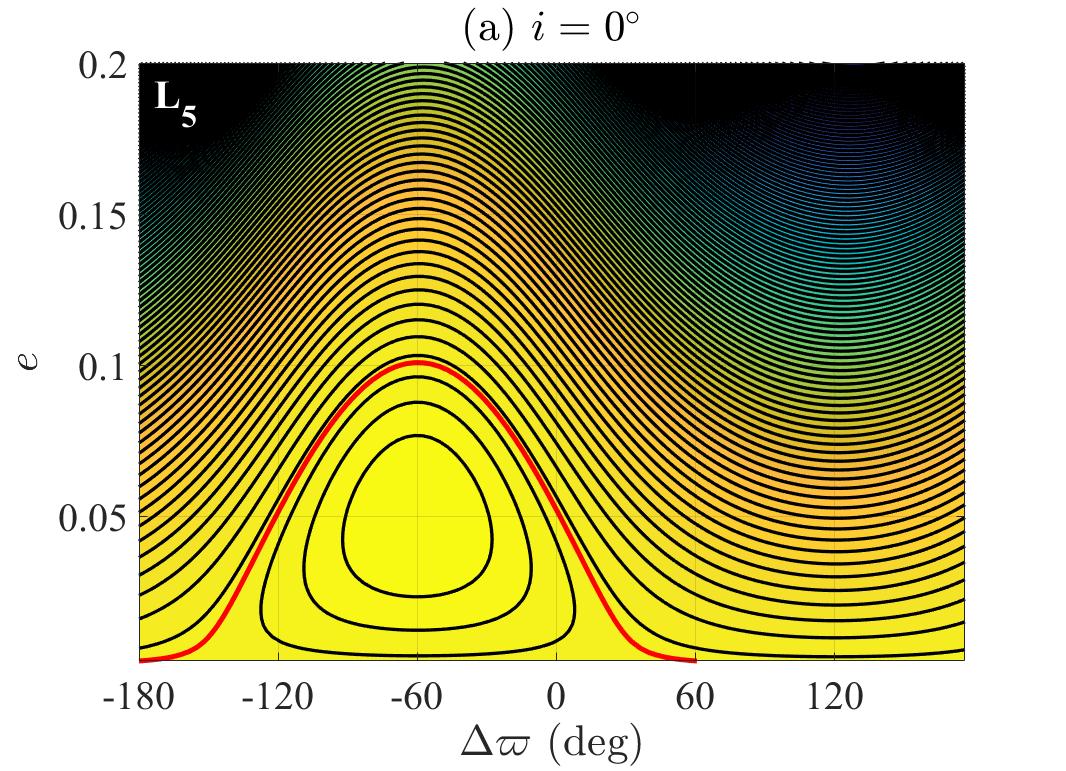}
  \end{minipage}
     \vspace{0 cm}
 \begin{minipage}[c]{1\textwidth}
  \vspace{0.2 cm}
  \includegraphics[width=9cm]{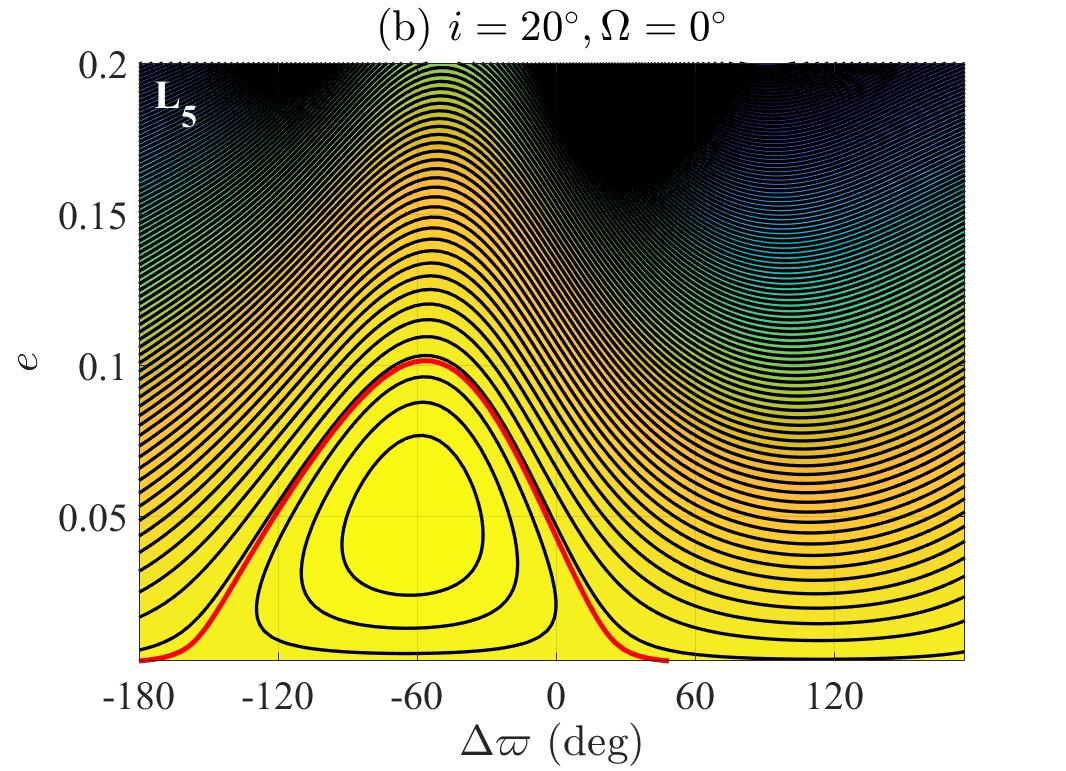}
  \includegraphics[width=9cm]{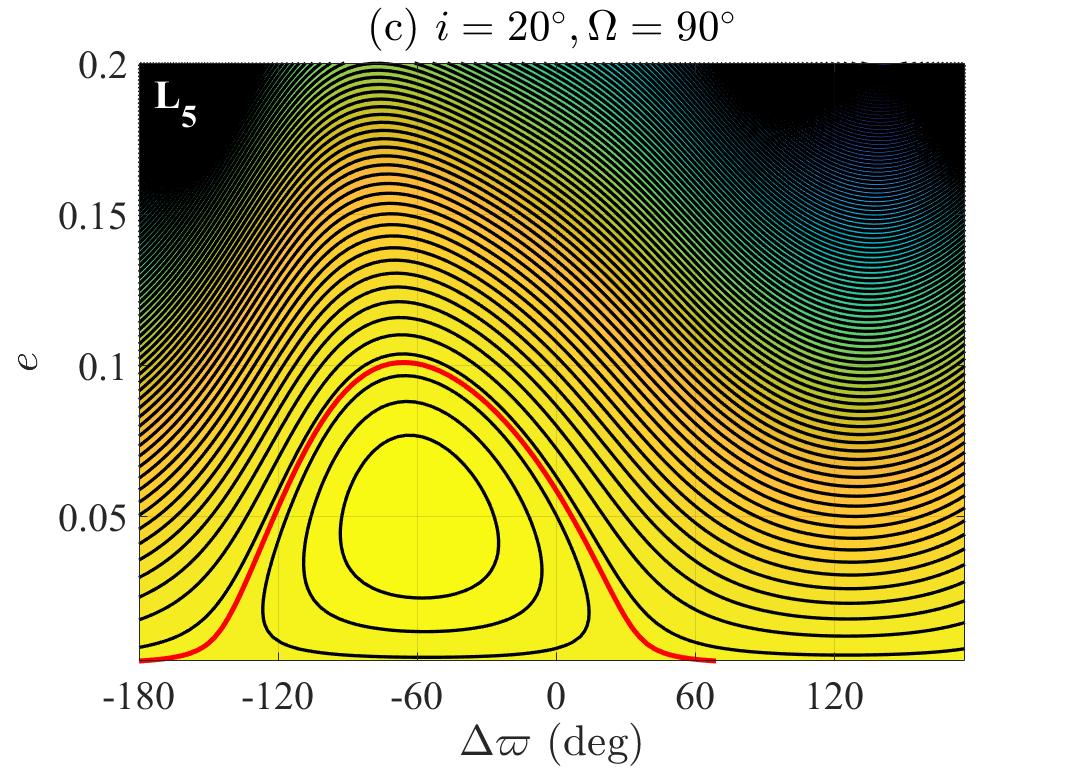}
  \end{minipage}
     \vspace{0 cm}
\begin{minipage}[c]{1\textwidth}
  \vspace{0.2 cm}
  \includegraphics[width=9cm]{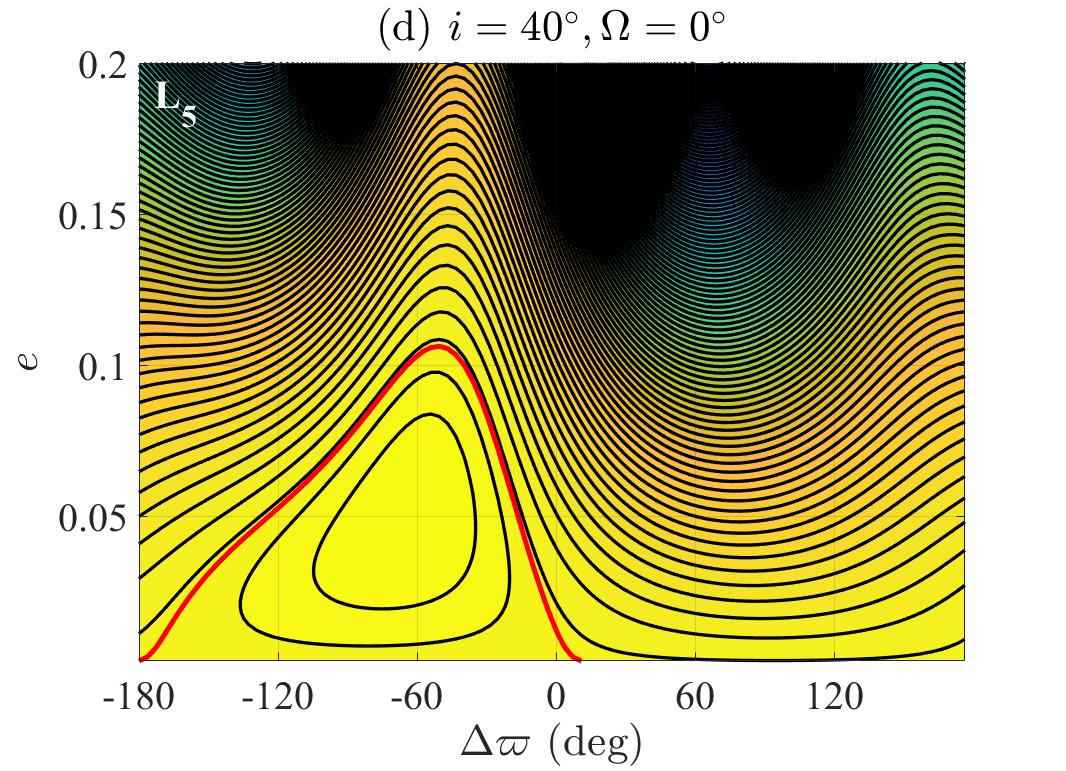}
  \includegraphics[width=9cm]{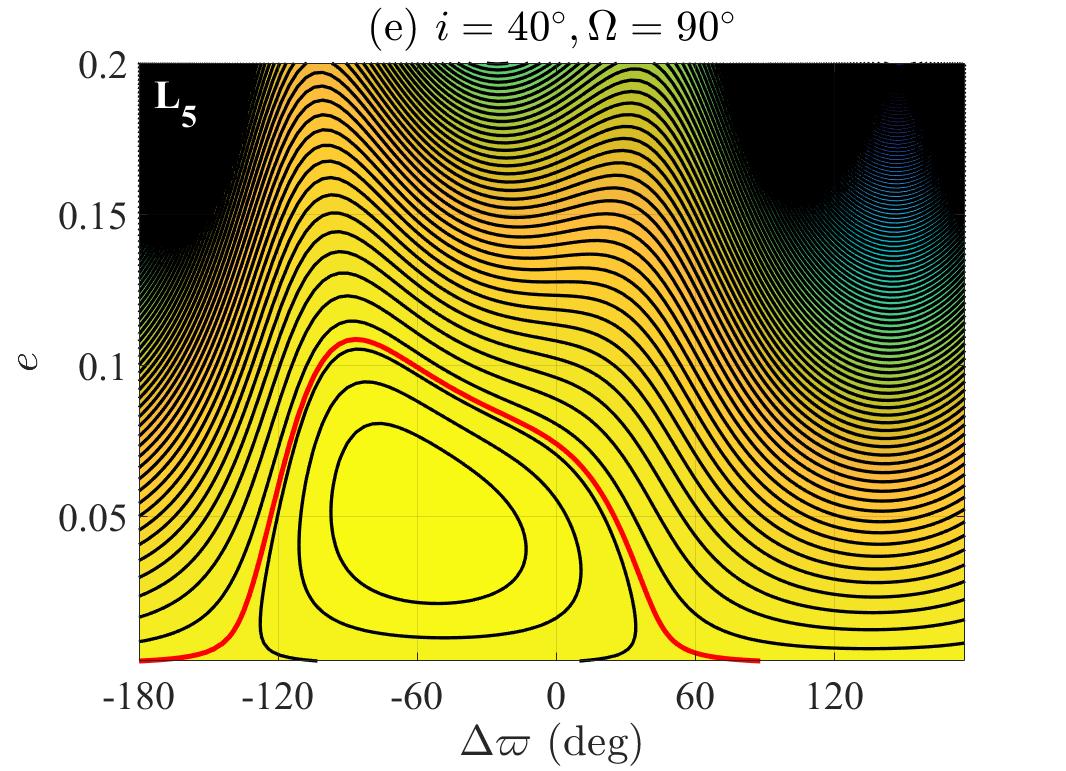}
  \end{minipage}
     \vspace{0 cm}
  \caption{The same as Fig. \ref{theory1}, but for the Lagrangian point L5. It shows that the stable equilibrium point locates around $\Delta\varpi=\varpi-\varpi_{\mbox{\scriptsize{J}}}=-60^{\circ}$ for the L5 Trojan swarm.}
 \label{theory1b}
 \end{figure*}

In the framework of the SERTBP, the instantaneous Hamiltonian can be written as
\begin{equation}\label{Eq1}
\mathcal{H}=-\frac{\mu}{2a}+n_J \Lambda_{\mbox{\scriptsize{J}}}-\mathcal{R}(a,e, i,\Delta\varpi, \Omega, \lambda,\lambda_{\mbox{\scriptsize{J}}}), ~~~~~~\mu=Gm_{\odot},
\end{equation}
where $G$ is the gravitational constant, $m_{\odot}$ is the mass of the Sun, $n_{\mbox{\scriptsize{J}}}$ is the mean motion of Jupiter, and $\Lambda_{\mbox{\scriptsize{J}}}$ is the action variable conjugated to Jupiter's mean longitude $\lambda_{\mbox{\scriptsize{J}}}$. The disturbing function $\mathcal{R}$ arising in equation (\ref{Eq1}) is related to the semimajor axis $a$, eccentricity $e$, inclination $i$ and mean longitude $\lambda$ of the Jupiter Trojan, together with the relative longitude of pericenter $\Delta\varpi ~(=\varpi - \varpi_{\mbox{\scriptsize{J}}})$, the longitude of ascending node $\Omega$ and Jupiter's mean longitude  $\lambda_{\mbox{\scriptsize{J}}}$ (see \citet{morb02} or \citet{murr99} for more details).

To discuss the resonance dynamics of Jupiter Trojans, let us introduce the critical argument
\begin{equation}\label{Eq2}
\psi = \lambda - \lambda_{\mbox{\scriptsize{J}}}.
\end{equation}
Thus, the disturbing function can be denoted by ${\cal R} (a, e, i,\Delta\varpi, \Omega,\psi,\lambda_{\mbox{\scriptsize{J}}})$. Since the Trojans are trapped in the 1:1 MMR, the resonant angle $\psi$ has a slow time evolution.  Therefore, we can average the disturbing function over the fast variable $\lambda_{\mbox{\scriptsize{J}}}$ to remove it
\begin{equation}\label{Eq3}
{{\cal R}^{\rm{*}}}(a, e, i,\Delta\varpi, \Omega,\psi ) = \frac{1}{{2\pi }}\int\limits_0^{2\pi } {{\cal R}~\mbox{d}{\lambda _{\mbox{\scriptsize{J}}}}}.
\end{equation}

For convenience, we choose the modified Delaunay's variables \citep{morb02},
\begin{equation}\label{Eq4}
\begin{aligned}
&\lambda=M+ \varpi,~~~~~~~~~~~\Lambda  = \sqrt {\mu a}, \\
& p =  - \varpi ,~~~~~~~~~~~~~~~P = \sqrt {\mu a} \left( {1 - \sqrt {1 - {e^2}} } \right),\\
& q = - \Omega, ~~~~~~~~~~~~~~~~ Q=\sqrt {\mu a} \left( {1 - \sqrt {1 - {e^2}} } \right)(1-\cos i),\\
&\lambda_{\mbox{\scriptsize{J}}}, ~~~~~~~~~~~~~~~~~~~~~~~{\Lambda _{\mbox{\scriptsize{J}}}},\\
&\varpi_{\mbox{\scriptsize{J}}},~~~~~~~~~~~~~~~~~~~~~~P_{\mbox{\scriptsize{J}}},
\end{aligned}
\end{equation}
which can be canonically transformed to
\begin{equation}\label{Eq5}
\begin{aligned}
{\varphi _1} &= \lambda  - {\lambda _{\mbox{\scriptsize{J}}}} = \psi ,~~~~~~~~~~~~~~~\quad {\Phi _1} = \Lambda, \\
{\varphi _2} &=  - p - {\varpi _{\mbox{\scriptsize{J}}}} = \Delta \varpi ,~~~~~~~~~\quad {\Phi _2} =  - P,\\
{\varphi _3} &= -q=\Omega,~~~~~~~~~~~~~~~~~~~~\quad {\Phi _3} = -Q,\\
{\varphi _4} &= {\lambda _{\mbox{\scriptsize{J}}}},~~~~~~~~~~~~~~~~~~~~~~~~~~~\quad {\Phi _4} = {\Lambda _{\mbox{\scriptsize{J}}}} + \Lambda,\\
{\varphi _5} &= {\varpi _{\mbox{\scriptsize{J}}}},~~~~~~~~~~~~~~~~~~~~~~~~~~~\quad {\Phi _5} = {P _{\mbox{\scriptsize{J}}}} - P,
\end{aligned}
\end{equation}
through the following generating function,
\begin{equation}\label{Eq6}
{\cal S} = \left( {\lambda  - {\lambda _{\mbox{\scriptsize{J}}}}} \right){\Phi _1} - \left( {p + {\varpi _{\mbox{\scriptsize{J}}}}} \right){\Phi _2} - q{\Phi _3} + {\lambda _{\mbox{\scriptsize{J}}}}{\Phi _4} + {\varpi _{\mbox{\scriptsize{J}}}} {\Phi _5}.
\end{equation}
Under the new set of canonical variables, the averaged Hamiltonian becomes
\begin{equation}\label{Eq7}
\begin{aligned}
{{\cal H}^{{*}}}& =  - \frac{{{\mu ^2}}}{{2\Phi _1^2}} - n_{\mbox{\scriptsize{J}}} {\Phi _1} - {{\cal R}^{\rm{*}}}\left( {{\Phi _1},{\Phi _2},{\Phi _3}, \psi, \Delta \varpi, \Omega } \right),\\
&=  - \frac{{{\mu ^2}}}{{2a}} - n_{\mbox{\scriptsize{J}}}\sqrt {\mu a} - {{\cal R}^{\rm{*}}}\left( {a, e, i, \psi, \Delta \varpi, \Omega } \right)
\end{aligned}
\end{equation}
Since the longitude of perihelion of Jupiter does not change in the SERTBP, the value of ${\varpi _{\mbox{\scriptsize{J}}}}$ could be set to be 0. Then both the variables ${\varpi _{\mbox{\scriptsize{J}}}}$ and  $\lambda_{\mbox{\scriptsize{J}}}$ are cyclic coordinates, and equation (\ref{Eq7}) actually determines a dynamical model with three degrees of freedom, i.e., $(\varphi _1, \varphi _2, \varphi _3, \Phi _1, \Phi _2, \Phi _3)$.

We assume that the critical argument $\psi$ is fixed at the classical Lagrangian point, i.e., $\psi = \lambda - \lambda_{\mbox{\scriptsize{J}}} = 60^{\circ}$ for L4 or $\psi = \lambda - \lambda_{\mbox{\scriptsize{J}}} = -60^{\circ}$ for L5. Under such a typical assumption \citep{nesv02b}, the dynamical model determined by equation (\ref{Eq7}) reduces to a system of two degrees of freedom, in which $\Delta \varpi$ and $\Omega$ are the associated angular coordinates. As a consequence, for a given pair of ($i$, $\Omega$), the global dynamical behaviors of Jupiter Trojans can be displayed by the level curves of the Hamiltonian in the pseudo phase space ($e$, $\Delta \varpi$).

Fig. \ref{theory1} presents the phase space structure in ($e$, $\Delta \varpi$) near Jupiter's 1:1 MMR, for the Lagrangian point L4. Corresponding to the planar case (i.e., $i=0$), Fig. \ref{theory1}a shows that there exists a stable equilibrium point at $\Delta \varpi \approx 60^{\circ}$. The apsidally aligned libration islands found here are consistent with the results from \citet{nesv02b}, and this could validate the Hamiltonian model adopted in this paper. Then we further consider the spatial co-orbital motion, the inclinations are fixed at large values of $i=20^{\circ}$ and $40^{\circ}$, given two representative longitudes of ascending nodes ($\Omega=0$ and $90^{\circ}$). The resulting portraits (Figs. \ref{theory1} b-d) show that, even in the highly inclined cases, the equilibrium point of $\Delta \varpi$ still exists around $60^{\circ}$; while the shapes of the libration islands may twist. Actually, according to our extra tests, the existence of such an equilibrium point is independent on the orbital elements $i\le40^{\circ}$ and $\Omega\in[0, 360^{\circ}]$. Taken in total, the theoretical approach indicates that the L4 Trojans can keep their orbital ellipses oriented with $\varpi\approx\varpi_{\mbox{\scriptsize{J}}}+60^{\circ}$, leading to the apsidal asymmetric-alignment depicted in Figs. \ref{Obs} and \ref{ObsMyr}.

We notice that the median eccentricity $e_m$ in the libration island is about 0.05, which likely corresponds to Jupiter's eccentricity ($e_{\mbox{\scriptsize{J}}}=0.05$). As a matter of fact, \citet{nesv02} showed that, in the planar case, $e_m$ is always approximately equal to $e_{\mbox{\scriptsize{J}}}$ as the latter artificially increases. This may qualitatively show that Jupiter's eccentric orbit does control the apsidal configuration of its Trojan population. Then we have to point out that, as long as the real value of $e_{\mbox{\scriptsize{J}}}=0.05$ is adopted, the libration islands locates below the separatrix going through a critical eccentricity of $e_c\approx0.1$ (see the red curve in Fig. \ref{theory1}a). This suggests that the Jupiter Trojans trapped in the apsidally aligned libration islands should have the maximum $e$ smaller than 0.1, otherwise $\Delta\varpi$ may experience a full circulation between 0 and $360^{\circ}$. While we find that the condition $e<e_c\approx0.1$ can only be fulfilled for a small fraction ($\sim20\%$) of the observed Jupiter Trojans, and some objects with current osculating $e<0.1$ may cross this upper limit during the long-term evolution. Besides, the inclinations of the Trojans are evolving with time but not steady, this could somewhat affect the profiles of the level curves on the phase space ($e$, $\Delta \varpi$) as indicated by Figs. \ref{Obs} and \ref{ObsMyr}. Therefore, in order to quantitatively understand the apsidal alignment phenomena, further investigation is clearly warranted.

For the Lagrangian point L5, the corresponding phase space structures at different $i$ values have also been constructed, as shown in Fig. \ref{theory1b}. It can been seen that, by comparing with Fig. \ref{theory1} (for the L4 point), these two sets of portraits show symmetry concerning the libration islands: (1) the stable equilibrium points are around $\Delta \varpi \approx -60^{\circ}$, where the L5 Trojans are clustered as shown in the right panels of Figs.\ref{Obs} and \ref{ObsMyr}; (2) the other properties of level curves, either in the libration islands or in the circulation regions, are nearly the same. This is not unexpected since the dynamics around the L5  point is a mirror image of the L4 point in the restricted three-body problem. Furthermore, as noted in Section 1 and some other previous works \citep{nesv02b, disi14, holt20}, under the gravitational effects of the four outer planets, the L4 and L5 Trojan swarms show almost the identical dynamical structure and stability in the long-term evolution. We thereby will only consider the L4 Jupiter Trojans in the following sections.

\subsection{Towards the explanation of the observation}

From the above semi-analytical results, we deduce that the apsidal alignment should appear for the Jupiter Trojans with $e<0.1$. When considering the dynamics of the Jovian Trojans, generally the proper elements are adopted \citep{mila93, beau01}. The proper elements can be understood as the average motion by eliminating the short periodic oscillations from the osculating elements \citep{morb05}. Following \citet{morb99} and \citet{lyka05},  we compute the \textit{numerical} proper eccentricities $e_p$ by averaging the 2000 yr time-step output from the entire 1 Myr integration\footnote{The orbital data used in this subsection is from the numerical simulations performed in Section 2, i.e., using the SWIFT\_RMVS3 symplectic integrator to calculate the dynamical evolution of the Trojans under the perturbations of four outer planets.}. We then find that, for the observed Trojans having $e_p<0.1$, a considerable portion of them actually are not confined by Jupiter's perihelion. We suppose that their osculating eccentricities may not be exclusively small  enough (i.e., $<0.1$), and the corresponding orbits could temporarily escape the $\Delta\varpi$ libration islands as shown in Fig. \ref{theory1}.

 \begin{figure}
  \centering
  \begin{minipage}[c]{0.5\textwidth}
  \centering
  \hspace{0cm}
  \includegraphics[width=8.2cm]{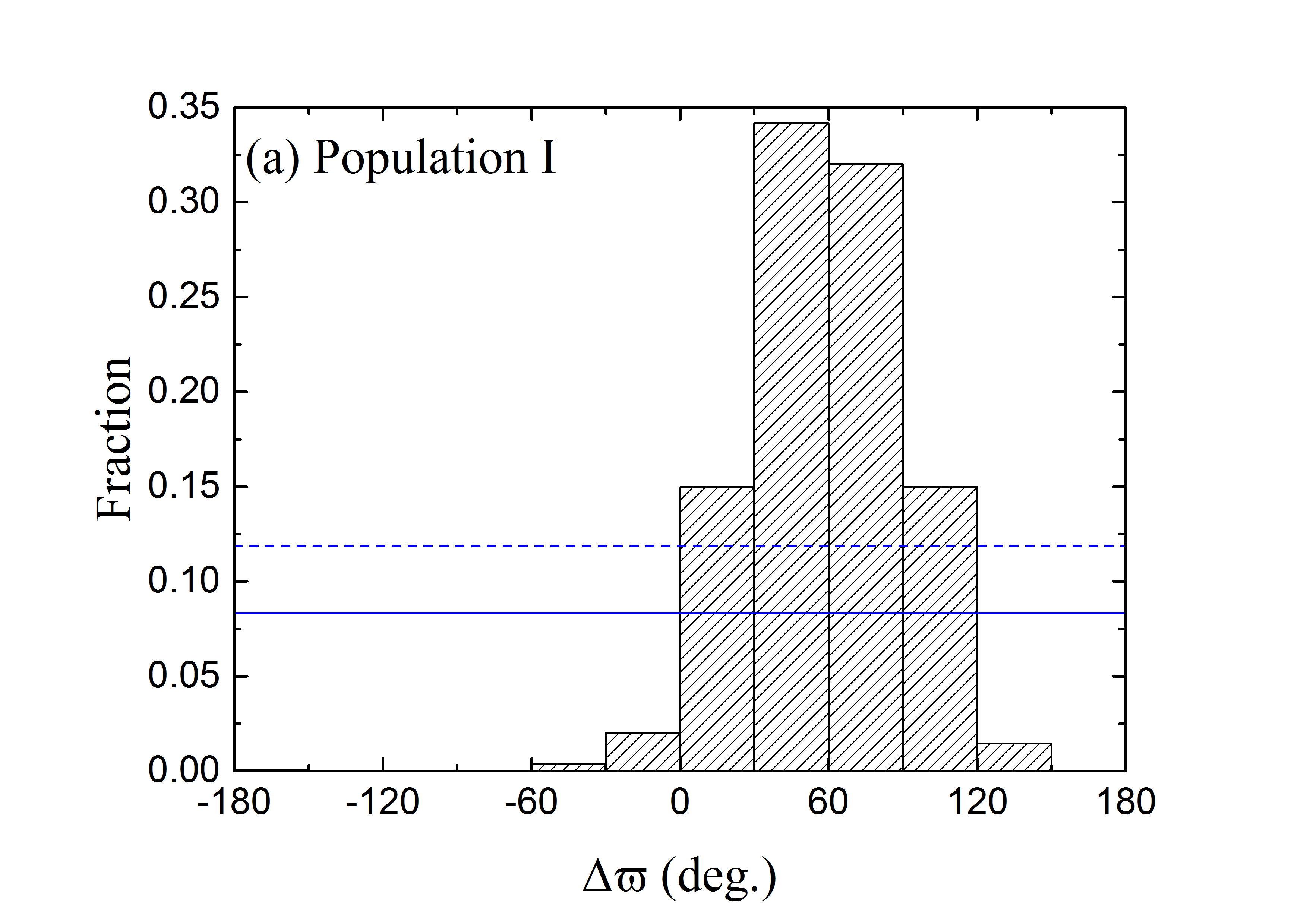}
  \end{minipage}
  \begin{minipage}[c]{0.5\textwidth}
  \centering
  \vspace{-0.1cm}
  \includegraphics[width=8.2cm]{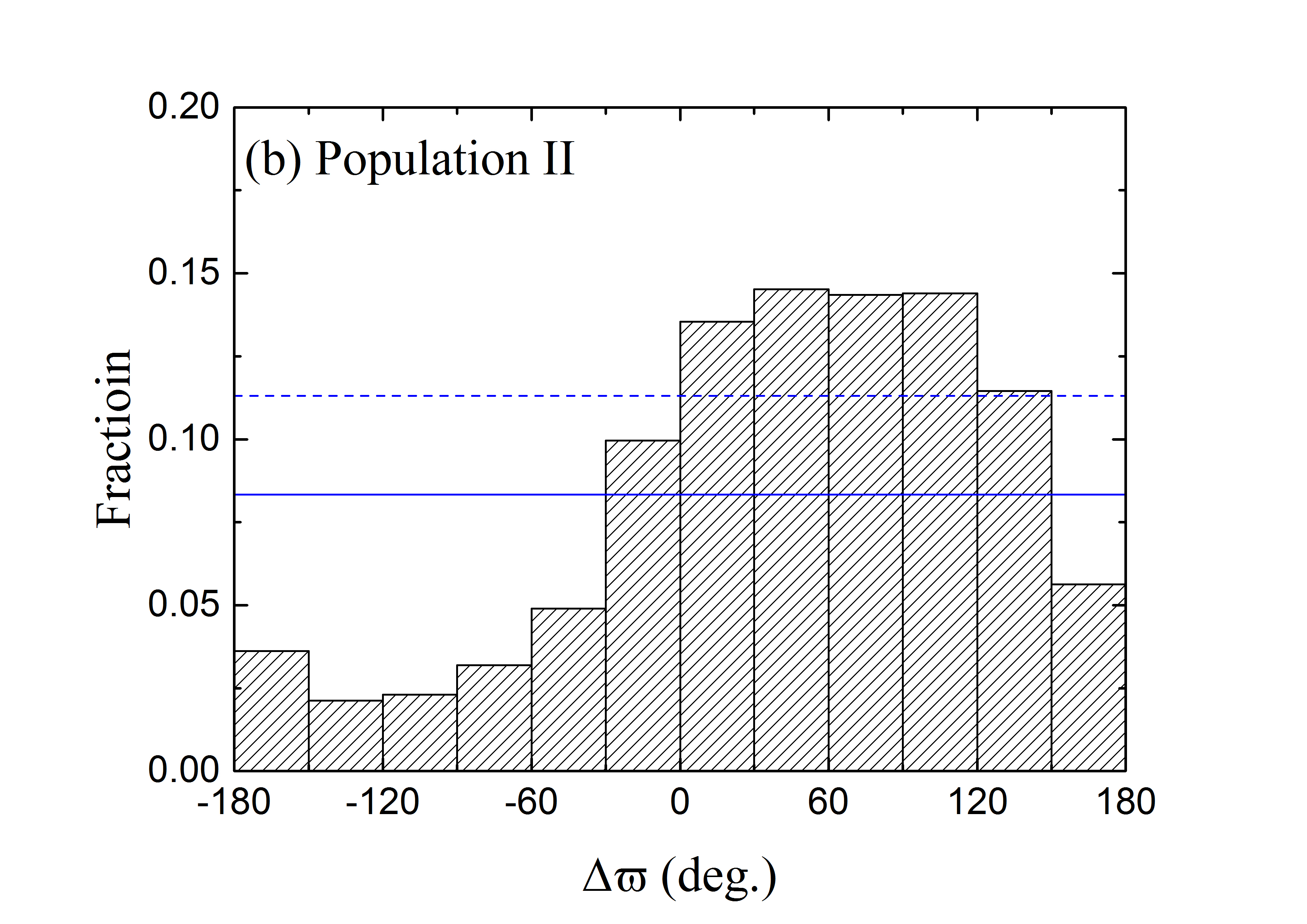}
  \end{minipage}
  \begin{minipage}[c]{0.5\textwidth}
  \centering
  \vspace{-0.1cm}
  \includegraphics[width=8.2cm]{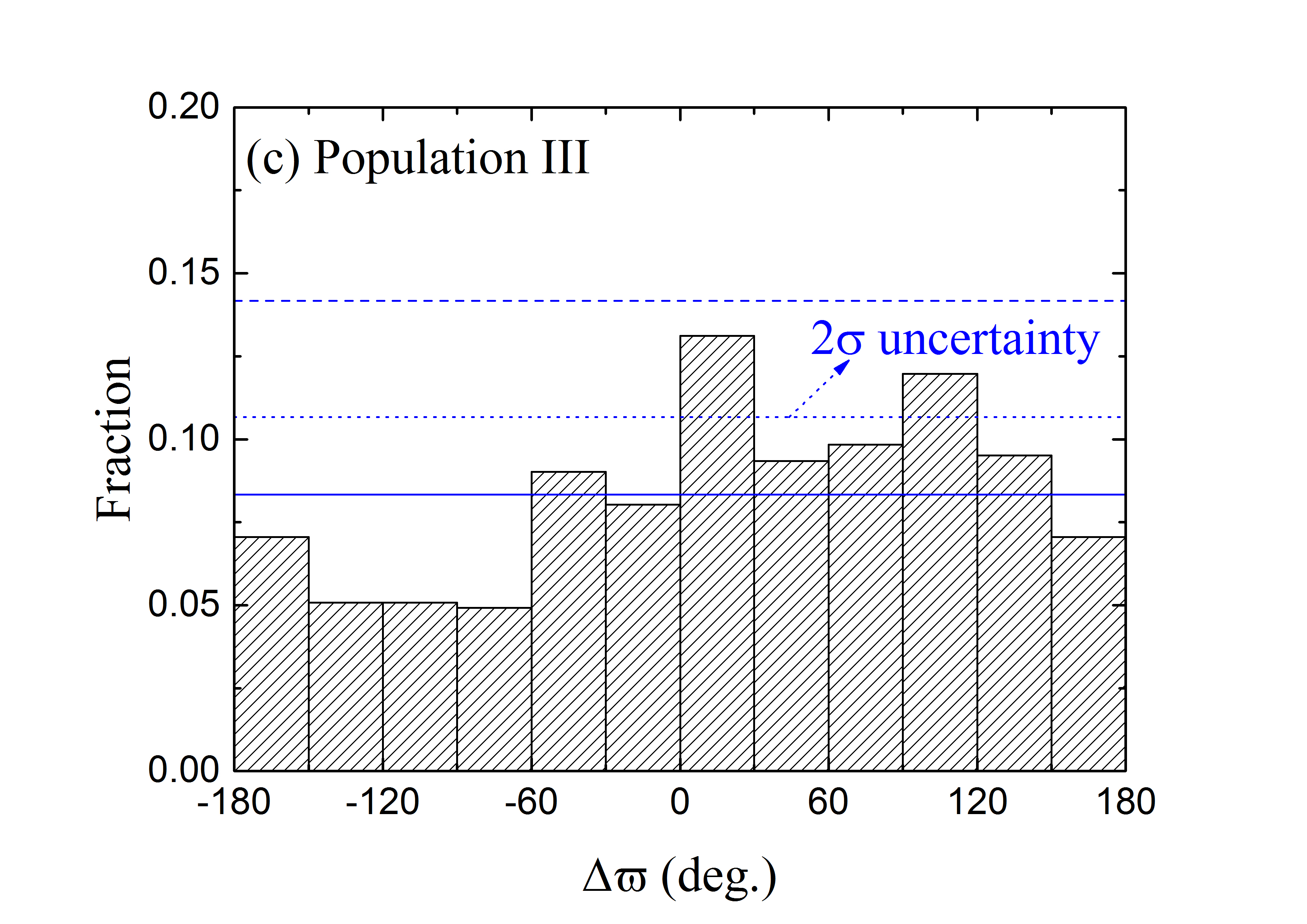}
  \end{minipage}
 \caption{Distribution of the difference of the longitudes of perihelia ($\Delta\varpi=\varpi-\varpi_{\mbox{\scriptsize{J}}}$) between a subset of L4 Trojans and Jupiter: (a) Population \Rmnum1 with maximum eccentricities $e_{\mbox{\scriptsize{Max}}}<0.1$ during the 1 Myr evolution; (b) Population \Rmnum2 with $e_{\mbox{\scriptsize{Max}}}>0.1$ but proper eccentricities $e_p<0.1$; (c) Population \Rmnum3 with $e_p>0.1$. The number fraction of a uniform $\Delta\varpi$ distribution, and its 5$\sigma$ upper uncertainty are also indicated by the horizontal solid and dashed lines, respectively. In the bottom panel, the 2$\sigma$ uncertainty is additionally plotted as the apsidal clustering is weaker for Population \Rmnum3.}
 \label{p123}
\end{figure}

In order to characterize the variation of the instantaneous osculating eccentricity, besides the proper value $e_p$ for eccentricity, we additionally introduce the maximum value $e_{\mbox{\scriptsize{Max}}}$. As we mentioned before, hereafter the analysis would be carried out only for the Jupiter Trojans around the L4 point. For the 4625 known L4 Trojans, we split them into three groups according to their eccentricities, and the individual distributions of $\Delta\varpi~(=\varpi-\varpi_{\mbox{\scriptsize{J}}})$ at the end of the 1 Myr integration are presented in Fig. \ref {p123}:

(1) \underline{Population \Rmnum1:} these objects have $e_{\mbox{\scriptsize{Max}}}<0.1$ during the entire 1 Myr evolution, surely the condition $e_p<0.1$ is fulfilled. The total number of this population is $N_1=1656$, i.e., a fraction of $p_1\sim36\%$ of the L4 Trojans. Since their eccentricities would never exceed 0.1, theoretically they are allowed to persistently reside in the islands of $\Delta\varpi$ libration shown in  Fig. \ref{theory1}.  Consequently, we notice in Fig. \ref {p123}a that, almost all of the Population \Rmnum1 samples are clustered in the interval $\Delta\varpi=0-120^{\circ}$, and the peak matches the theoretical equilibrium point at $\Delta\varpi=60^{\circ}$.

(2) \underline{Population \Rmnum2:} these objects have $e_p<0.1$ but $e_{\mbox{\scriptsize{Max}}}>0.1$ ($N_2=2348$ and $p_2\sim51\%$). This indicates that they would have their osculating eccentricities occasionally increased to $e>0.1$, while for a larger time fraction of the evolution their orbits lie within the $e<0.1$ range. Fig. \ref {p123}b shows that the apsidal confinement of Population \Rmnum2 is still quite noticeable, as the histogram bins around $\Delta\varpi=60^{\circ}$ could be higher than the dashed line, which indicates a significant variation from a uniform $\Delta\varpi$ distribution (see the solid line) at the $5\sigma$ confidence level.
\\\noindent Nevertheless, Population \Rmnum2 has a much less prominent clustering of $\varpi$ than Population \Rmnum1, because during some time intervals they have osculating eccentricities $e>0.1$ and populate outside the apsidal alignment islands in Fig. \ref{theory1}. Through a deeper analysis of the orbital evolution of the Population \Rmnum2 samples in our numerical integration, we find that the objects located in the vicinity of $\Delta\varpi=60^{\circ}$ keep changing. That is to say the objects aligned in $\varpi$ at a certain time $t_1$ could be much different from those at another time $t_2$. For instance, we considered a subset of Population \Rmnum2 with $e_{\mbox{\scriptsize{Max}}}=0.11-0.12$, which has a total number of 462. From the beginning of the integration, there are 88 objects trapped in the region of $\Delta\varpi=60^{\circ}\pm20^{\circ}$ at $t=0.1$ Myr, and 105 such apsidally aligned objects are identified at $t=1$ Myr, but only 11 are the same ones.

(3) \underline{Population \Rmnum3:} they have larger $e_p>0.1$ ($N_3=612$ and $p_3\sim13\%$) and spend a considerable fraction of time outside the theoretical libration islands of $\Delta\varpi$. Naturally, the distribution of $\Delta\varpi$ shown in Fig. \ref {p123}c becomes rather flat by comparing with Fig. \ref {p123}b. Nevertheless, we cannot fail noticing that although the apsidal alignment of Population \Rmnum3 is weak, but still visible. We determined a lower significance limit at about $2\sigma$ uncertainty, as indicated by the dotted line. So why not their $\varpi$ could disperse well enough to be close to a uniform distribution (represented by the solid line)? The reason should be attributed to the fact that the objects with minimum eccentricities larger than 0.1, i.e., could hardly experience the $\Delta\varpi$ libration phase, are as few as $<1\%$ among Population \Rmnum3.\\

\begin{figure}
 \hspace{0 cm}
  \includegraphics[width=8.5cm]{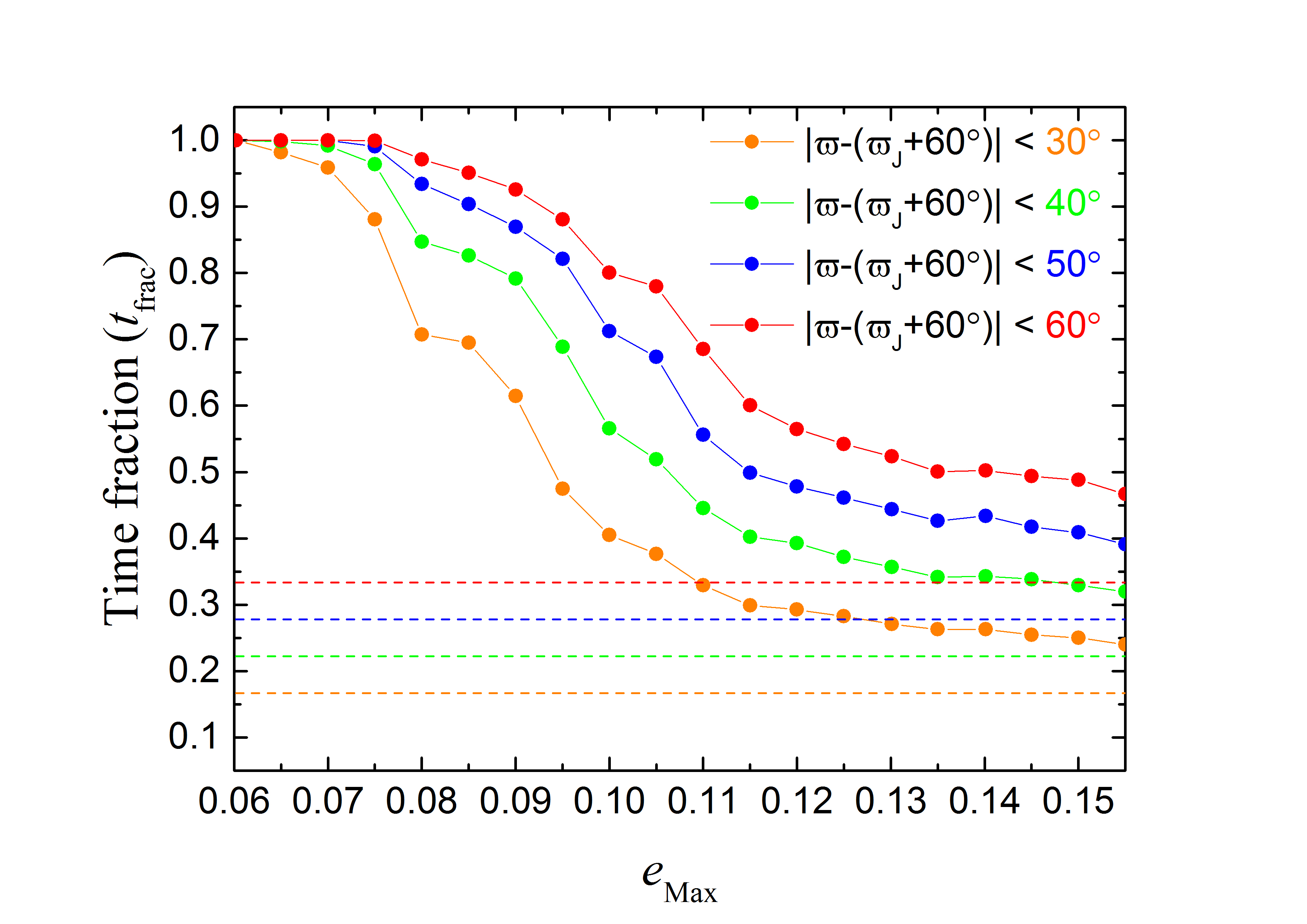}
  \caption{For the Jupiter Trojans with proper eccentricities $e_p<0.1$ from Population \Rmnum1 and Population \Rmnum2, as a function of their maximum eccentricities $e_{\mbox{\scriptsize{Max}}}$, the time fraction $T_{frac}$ that they spend in the range of $|\Delta\varpi-60^{\circ}|<\delta=30^{\circ}$ (orange), $40^{\circ}$ (green), $50^{\circ}$ (blue) and $60^{\circ}$ (red). The dashed line with the corresponding colour shows that, if the apsidal confinement does not emerge, the Trojans would have a fixed time fraction within the range of $|\Delta\varpi-\Delta\varpi_0|<\delta$ for any $\Delta\varpi_0\in[0, 360^{\circ}]$.}
  \label{time}
\end{figure}

In order to quantitatively measure how tight the apsidal clustering of Jupiter Trojans is,  similar to the study of the stickiness effect in chaotic system \citep{zhou14}, in this work we define a time index $T_{frac}$. During the 1 Myr evolution, once a Trojan settles into the neighborhood of the $\Delta\varpi$ equilibrium point, characterized by $|\Delta\varpi-60^{\circ}|<\delta$, the length of the time interval is recorded. Then the time index is computed as the total time fraction
\begin{equation}
T_{frac}=\frac{\sum_m\mbox{time interval}~(|\Delta\varpi-60^{\circ}|<\delta)}{\mbox{total integration time (=1 Myr)}},
\label{Tfrac}
\end{equation}
where $m$ is the number of times that the Trojan enters this $\Delta\varpi$ range. Now we consider the Jupiter Trojans with $e_p<0.1$, i.e., from Population \Rmnum1 and Population \Rmnum2. They comprise about $87\%$ of the L4 swarm and hence dominate the overall orbital distribution. Here $\delta$ is adopted to be $30^{\circ}$, $40^{\circ}$, $50^{\circ}$ and $60^{\circ}$, and correspondingly the variation of $T_{frac}$ is plotted as a function of $e_{\mbox{\scriptsize{Max}}}$ in Fig. \ref{time}. The dashed line indicates that, if an object spends equal time in the range of $|\Delta\varpi-\Delta\varpi_0|<\delta$ for any $\Delta\varpi_0\in[0, 360^{\circ}]$, the resultant time fraction is fixed at $T_{frac}=2\delta/360$.

The most important result depicted in Fig. \ref{time} is that, for these Trojans, the temporal variation in $\varpi$ is far from uniform. Taking the case of $\delta=60^{\circ}$ for example, the red curve always lies above the red dashed line, meaning that a Trojan with $e_p<0.1$ would have its $\varpi$ stuck in a narrow region centered on $\varpi_{\mbox{\scriptsize{J}}}+60^{\circ}$ for a considerable fraction of time. As a result, through a global view of the Trojan population at any specific moment, we would observe the apsidal clustering of their orbits. This phenomenon could be analogous to the spiral arms in galaxies, which is not merely a static accumulation of stars and dust, as explained by the density wave model \citep{lin69}. Especially, for the Population \Rmnum1 samples with $e_{\mbox{\scriptsize{Max}}}<0.1$ (i.e., the condition $e<e_c\approx0.1$ is always satisfied), they could be trapped in the range of $\Delta\varpi=60^{\circ}\pm60^{\circ}=0-120^{\circ}$ over a time fraction of $T_{frac}>80\%$. Such a large $T_{frac}$ can nicely explain the prominent components of Population \Rmnum1 confined in this very $\Delta\varpi$ range (see Fig. \ref{p123}a).

Additionally, the profiles of the curves in Fig. \ref{time} show that, $T_{frac}$ decreases as a function of increasing $e_{\mbox{\scriptsize{Max}}}$. Since the osculating eccentricity of a Trojan's orbit evolves with time, the maximum value $e_{\mbox{\scriptsize{Max}}}$ can represent a particular level curve in the phase space ($e$, $\Delta \varpi$), as displayed in Fig. \ref{theory1}. When $e_{\mbox{\scriptsize{Max}}}$ becomes larger, the libration island centered on the equilibrium point at $\Delta\varpi=60^{\circ}$ will be increasing in size, leading to fewer part of it located inside a given $\Delta\varpi$ range. Correspondingly, for the orbit in the physical space, it would be confined in the region $|\Delta\varpi-60^{\circ}|<\delta$ for a smaller time fraction $T_{frac}$. And obviously, the value of $T_{frac}$ is increasing with the width of the variation in $\Delta\varpi$, as characterized by the parameter $\delta$.

In summary, the vast majority of the observed L4 Trojans have proper eccentricities $e_p$ smaller than 0.1, and they are evolving on the orbits with osculating eccentricities $e<0.1$ for a large period of time in which they could wander inside the libration islands of $\Delta\varpi$. This scenario could account for the apsidal confinement of the entire Trojan population.

\section{Apsidal clustering: reproduction}

The planetesimals could have been chaotically captured into Jupiter's co-orbital regions during the early evolution of the solar system within the Nice model \citep{tsig05, morb05}. At the end of this process, their longitudes of perihelia should cover all values of $\varpi=0-360^{\circ}$. Therefore, we would like to investigate that whether the apsidal asymmetric-alignment of the currently observed Jupiter Trojans, could be generated from an even $\varpi$ distribution in the later long-term evolution. As the study in the previous sections is indicative of this peculiar orbital structure, the numerical experimentation is to be carried out to evaluate the effect of the secular perturbation of the eccentric Jupiter.

 \begin{figure}
  \centering
  \begin{minipage}[c]{0.5\textwidth}
  \centering
  \hspace{0cm}
  \includegraphics[width=8.5cm]{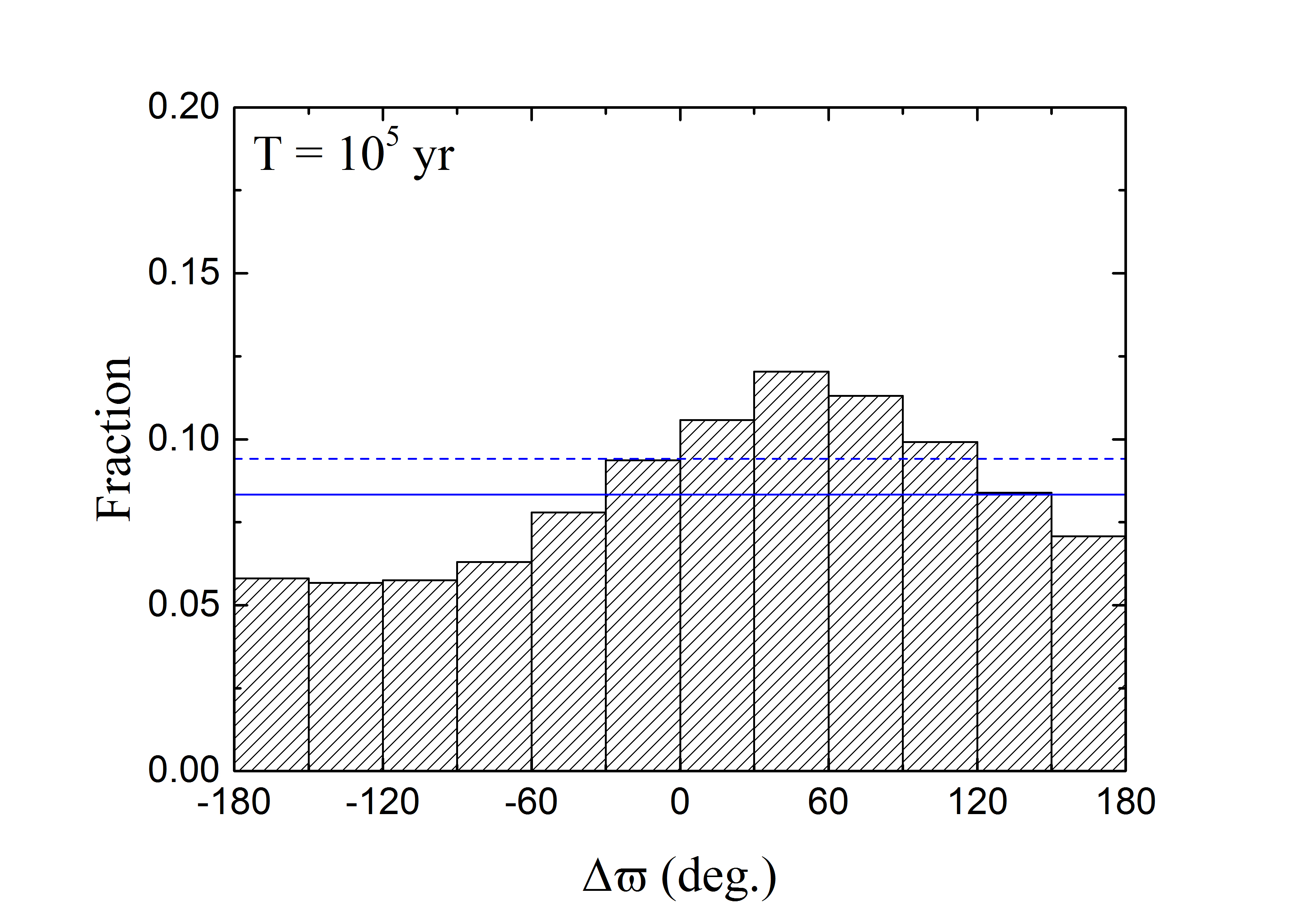}
  \end{minipage}
  \begin{minipage}[c]{0.5\textwidth}
  \centering
  \vspace{0cm}
  \includegraphics[width=8.5cm]{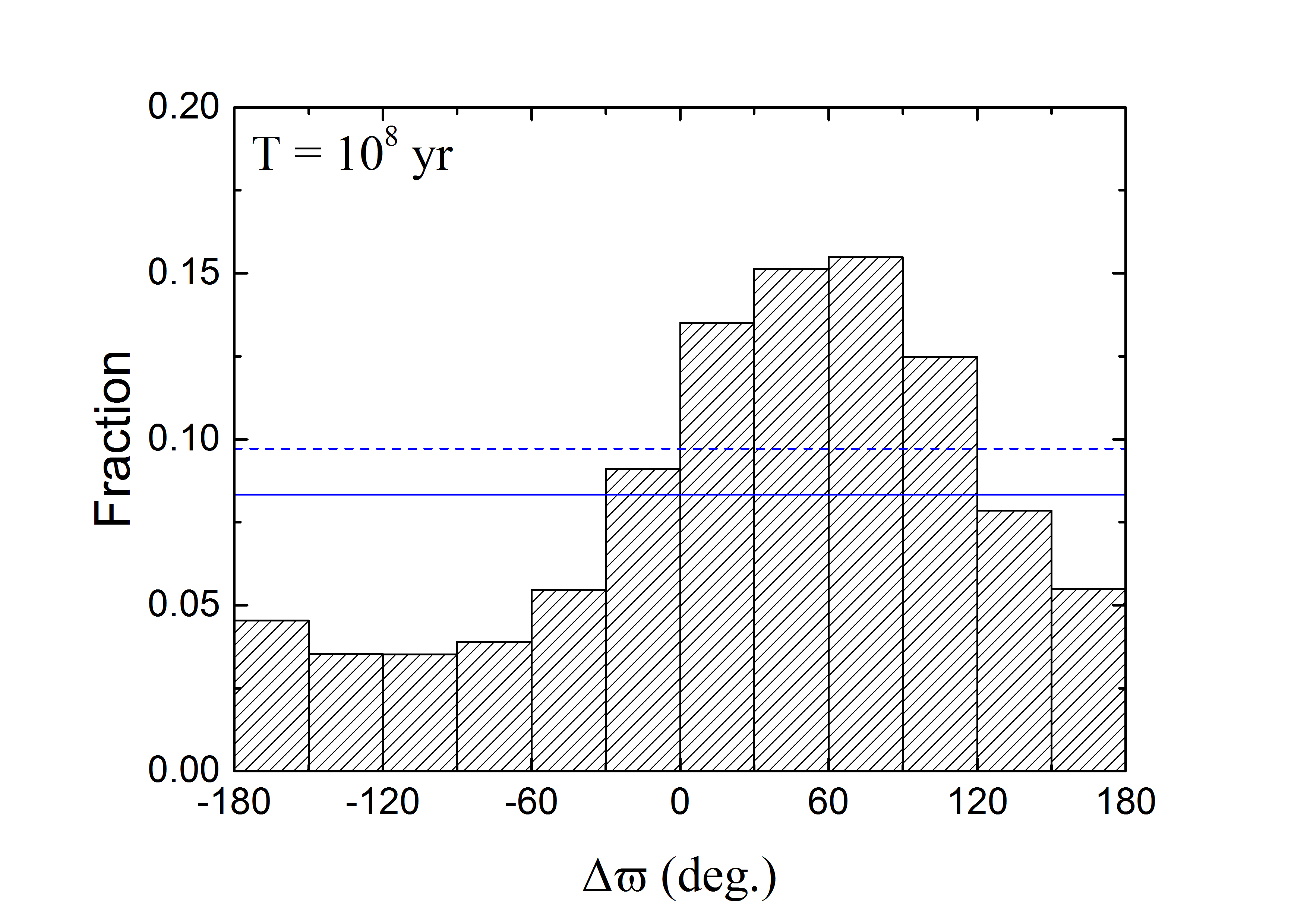}
  \end{minipage}
  \begin{minipage}[c]{0.5\textwidth}
  \centering
  \vspace{0cm}
  \includegraphics[width=8.5cm]{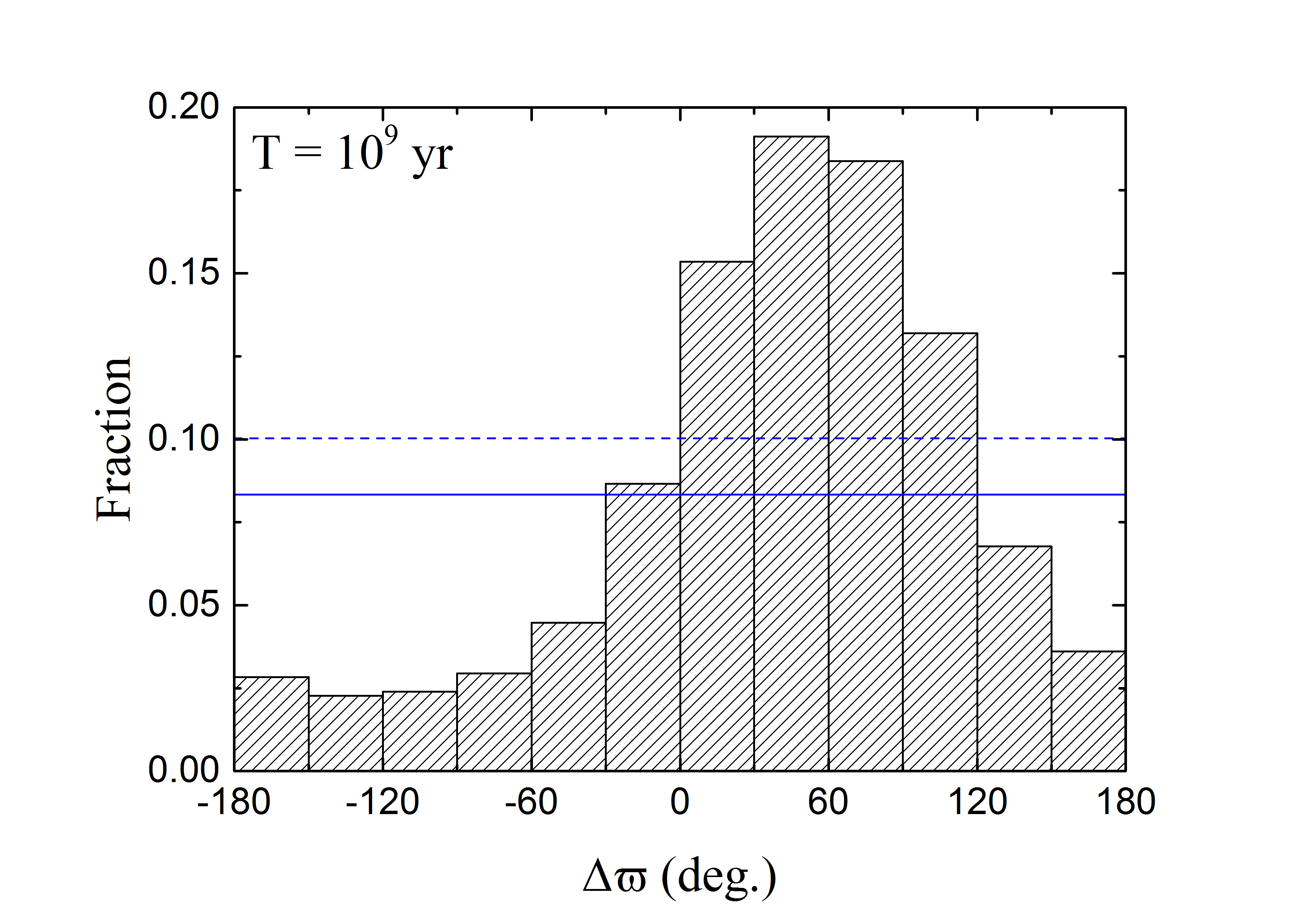}
  \end{minipage}
 \caption{For simulated Jupiter Trojans starting with longitudes of perihelia chosen uniformly in the range of $\varpi=0-360^{\circ}$ (equivalent to $\Delta\varpi=\varpi-\varpi_{\mbox{\scriptsize{J}}}=0-360^{\circ}$), the evolving distribution of $\Delta\varpi$ at $t=10^5$ yr (top panel), $10^8$ yr (middle panel) and $10^9$ yr (bottom panel). The solid and dashed lines represent the uniform distribution of $\Delta\varpi$ and the $5\sigma$ significance, respectively.}
 \label{tp}
\end{figure} 

 \begin{figure*}
  \centering
  \begin{minipage}[c]{1\textwidth}
  \vspace{0 cm}
  \includegraphics[width=16cm]{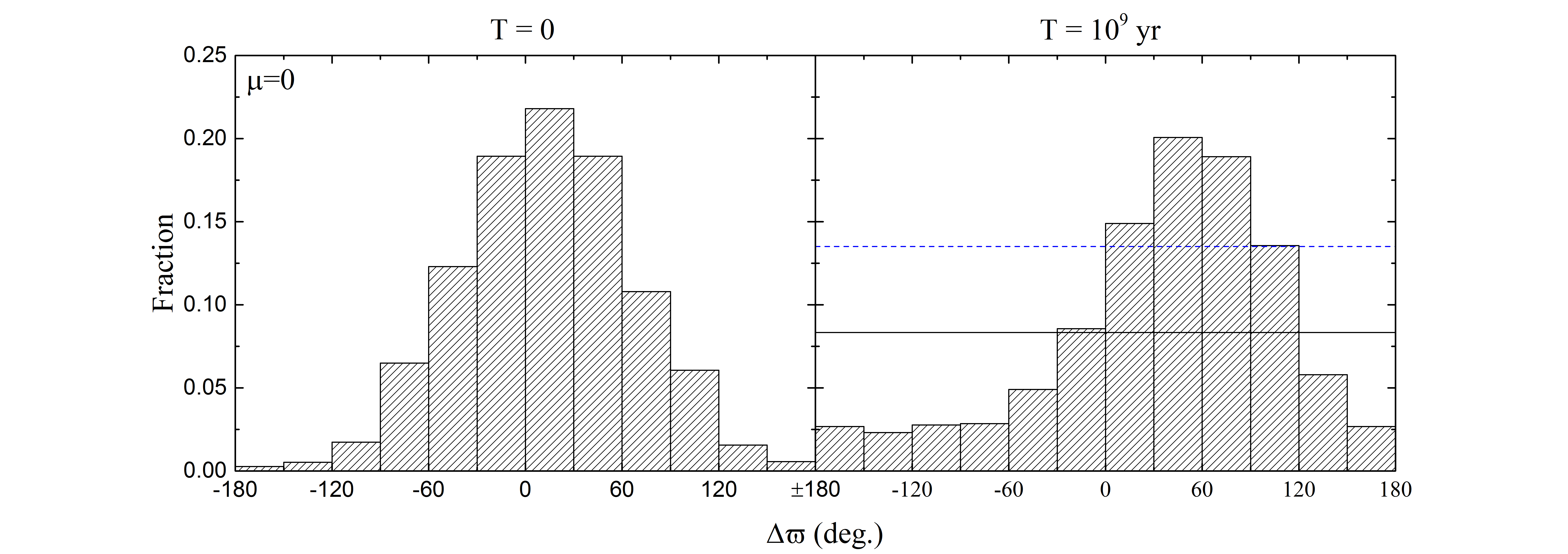}
  \end{minipage}
     \vspace{0 cm}
 \begin{minipage}[c]{1\textwidth}
  \vspace{0 cm}
  \includegraphics[width=16cm]{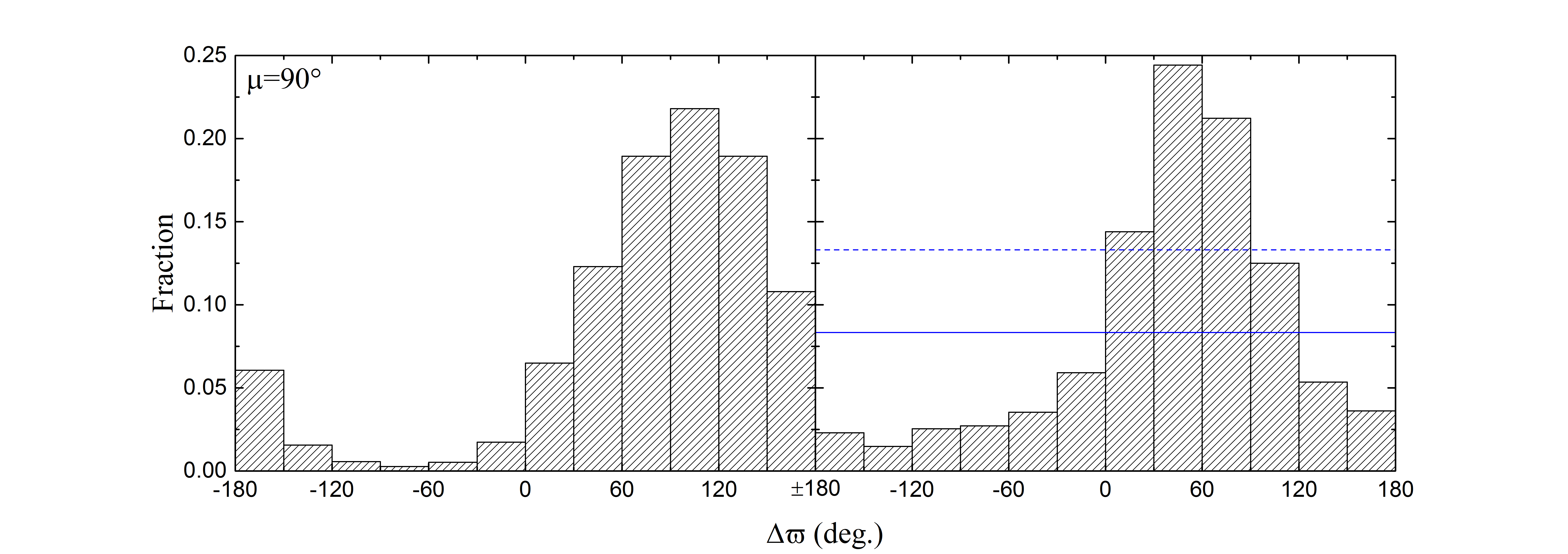}
  \end{minipage}
     \vspace{0 cm}
\begin{minipage}[c]{1\textwidth}
  \vspace{0 cm}
  \includegraphics[width=16cm]{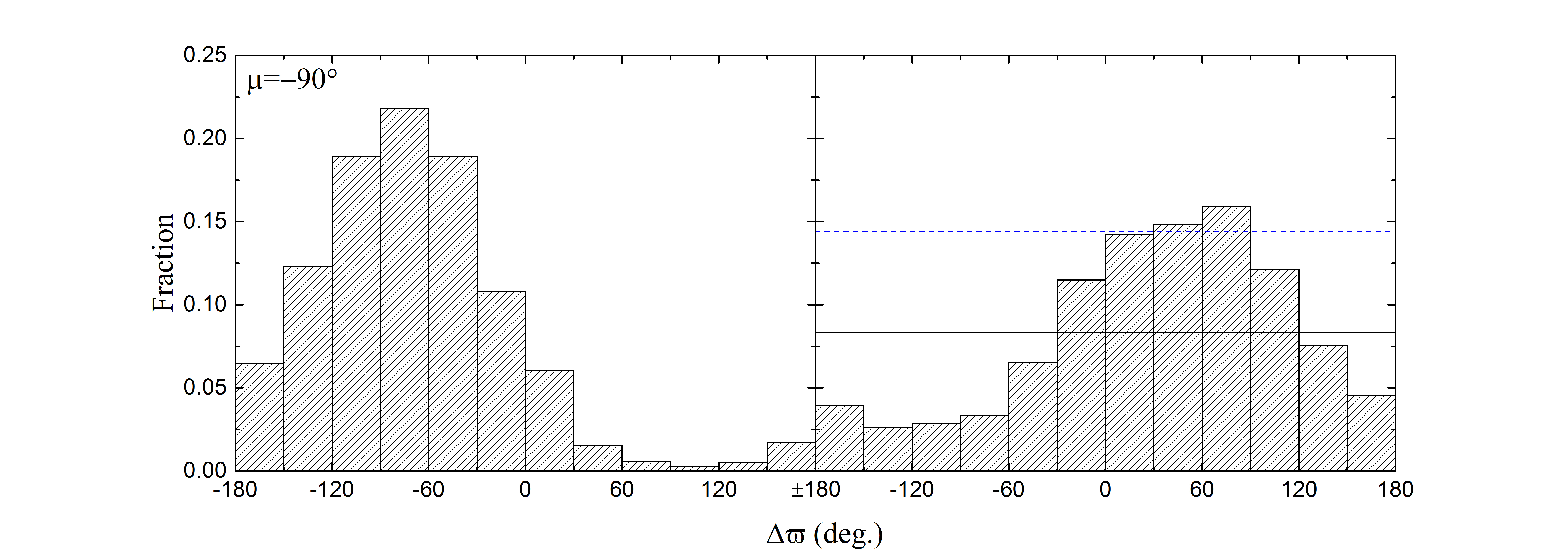}
  \end{minipage}
     \vspace{0 cm}
  \begin{minipage}[c]{1\textwidth}
  \vspace{0 cm}
  \includegraphics[width=16cm]{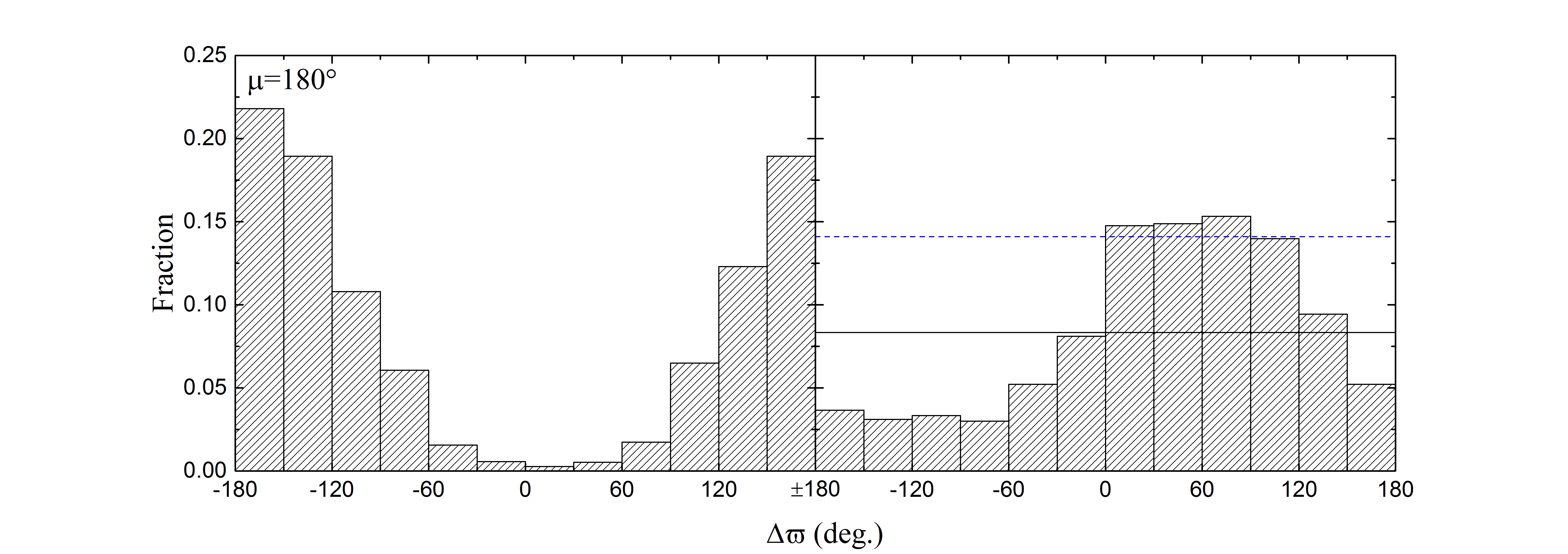}
  \end{minipage}
     \vspace{0 cm}
  \caption{\textit{Left column}: Initial $\Delta\varpi$ of simulated Jupiter Trojans assigning from Gaussian distributions with means of $\mu=0, 90^{\circ}, -90^{\circ}, 180^{\circ}$ (from top to bottom). For the bin next to the location $\Delta\varpi=\mu$, the initial number fraction could peak over 0.2.  \textit{Right column}: The corresponding final distribution of $\Delta\varpi$ after $10^9$ yr evolution. As well, the uniform distribution (solid line) and its $5\sigma$ significance (dashed line) are indicated in each panel.}
 \label{tpgauss}
 \end{figure*}

In this section, we still adopt the present outer solar system model, consisting of the Sun and four giant planets (i.e., Jupiter, Saturn, Uranus and Neptune). As in recent dynamical explorations of the Jupiter Trojans (e.g., \citet{hell19}, \citet{holt20}), to construct robust simulations, we follow the long-term evolution of 30,000 test Trojans around the L4 point. Initially, all particles have the same semimajor axis ($a\approx5.2$ AU) with Jupiter. And similar to the observed Trojan population, they have $e$ ranging from $0-0.3$ and a broad inclination distribution of $i=0-40^{\circ}$. The initial orbital angles are chosen such that particles start from $\Omega=0$ and $\psi=\lambda-\lambda_{\mbox{\scriptsize{J}}}=30^{\circ}-90^{\circ}$, while the values of their $\varpi$ are randomly selected between 0 and $360^{\circ}$. We then monitor the orbital evolution of these test Trojans in the long-term numerical integration. 

Bearing in mind that, for Jupiter, the periodicity in the variation of its longitude of perihelion $\varpi_{\mbox{\scriptsize{J}}}$ is about $1.7\times10^5$ yr. So given a relatively shorter timescale of $10^5$ yr, in the top panel of Fig. \ref{tp}, we provide the first glance at the evolving distribution of $\Delta\varpi=\varpi-\varpi_{\mbox{\scriptsize{J}}}$. At this moment, there are a total of 26,117 particles (i.e., $87.1\%$) found on Trojan orbits, characterized by the librating critical arguments $\psi$. We notice that, although the distribution of $\Delta\varpi$ is seemingly accumulated around $60^{\circ}$, it is in fact very close to the uniform distribution indicated by the solid line.

As the time passes, due to the secular perturbation of the eccentric Jupiter, the simulated Trojans can reach the orbits which are more and more apsidally aligned. At the epoch of $10^8$ yr, as shown in the middle panel of Fig. \ref{tp}, the clustering in $\Delta\varpi$ is quite obvious. And when the integration ends at $10^9$ yr, on the order of the age of the solar system, there remain 10,417 objects (i.e., $\sim34.7\%$ of initial samples) librating around the L4 point. For these simulated Trojans, as shown in the bottom panel of Fig. \ref{tp}, the largest fraction of objects clustered adjacent to the location of $\Delta\varpi=60^{\circ}$ is as high as $\sim0.2$. We recall that the distribution of the longitudes of perihelia of the observed L4 Trojans, as displayed in the left panels of Figs. \ref{Obs} and \ref {ObsMyr}, also has a comparable peak next to the same location. 

At the beginning of this section we assumed that the planetesimals deposited in Jupiter's Trojan clouds in the framework of the Nice model have uniformly-distributed $\varpi$. Although it is a reasonable assumption since the evolution of the planetesimal is chaotic, the capture into libration around the L4 or L5 point could be dependent on the pre-capture orbit of the planetesimal. For instance, such a dependence may be that the perihelion distance of the planetesimal is smaller than that of Jupiter. Besides, there also may be other constraints imposed on the planetesimal's orbit including the angle $\varpi$ describing its orientation. Then some additional simulations have been carried out, by assigning initial $\Delta\varpi$ values randomly from a Gaussian distribution 
\begin{equation}\label{gauss}
f(\Delta\varpi)=\frac{1}{\sqrt{2\pi}\hat{\sigma}}\mbox{exp}\left[-\frac{(\Delta\varpi-\mu)^2}{2{\hat{\sigma}}^2}\right],
\end{equation}
where $\mu$ is the mean,  and the standard deviation $\hat{\sigma}$ is taken to be 0.3. The initial $\varpi$ of test Trojans can simply determined as $\varpi_{\mbox{\scriptsize{J}}}+\Delta\varpi$. 

For each set of 3000 test Trojans, we choose the mean $\mu$ to be one of four representative values of $0, 90^{\circ}, -90^{\circ}, 180^{\circ}$, and the associated $\Delta\varpi$ distribution is plotted in the left column in Fig. \ref{tpgauss} (from top to down). We can see that the initial number fraction of test Trojans within the bin next to $\Delta\varpi=\mu$ could peak over 0.2. Keeping the other initial conditions of the system, we repeat the $10^9$ yr numerical integrations for the simulated Jupiter Trojans, and the final distributions of $\Delta\varpi$ are summarized in the right column in Fig. \ref{tpgauss}. The main feature is that, starting from any $\Delta\varpi$ distribution, the simulated Trojans would eventually reach the state of apsidal alignment around $\Delta\varpi=60^{\circ}$. Through a closer look at the cases of $\mu=-90^{\circ}, 180^{\circ}$, one notices that although the region near $\Delta\varpi=60^{\circ}$ was nearly empty (two lower left panels), the expected apsidal alignment of the Trojans within this region can still be seen at the end of the $10^9$ yr evolution (two lower right panels). This provides support for the results obtained from our previous run with an initially uniform $\Delta\varpi$ distribution. We end by noting that the previous run using 30000 test Trojans is quite ``expensive'', and more than a month of CPU time is needed. As a matter of fact, we found that an order of magnitude fewer objects could be a sufficient sample size to perform the statistical analysis.  Therefore, in order to save computational time, we use a set of 3000 test Trojans (for each $\mu$) for the simulations with different set-up of $\Delta\varpi$ distribution, and the results are considered robust.

%Through a closer look at the cases of $\mu=-90^{\circ}, 180^{\circ}$, one sees that the fractions of the Trojans clustered in the vicinity of $\Delta\varpi=60^{\circ}$ are relatively lower. This result is not surprising, because as the two lower left panels show, the regions near $\Delta\varpi=60^{\circ}$ are nearly empty due to the simulation set-up. Even so, the expected apsidal alignment around this $\Delta\varpi$ can still be obtained at the end of the $10^9$ evolution.

In summary, we conclude that the apsidal asymmetric-alignment of Jupiter Trojans should be a natural consequence of Jupiter's secular perturbation. The orbital orientations of the Jupiter Trojans at the moment of capture could affect the extent of the clustering in their $\varpi$, but only slightly and would not alter the overall $\varpi$ distribution. The origin of the Jupiter Trojans still remains uncertain, and a thorough exploration goes beyond the scope of this paper.

%________________________________________________________________________________________________________________________________________________________

\section{Conclusions and discussion}

This work is inspired by the hypothesis that the eccentric and inclined Planet 9 could cause the orbital clustering of the extreme KBOs beyond 250 AU \citep{baty16}. As Jupiter has a relatively large eccentricity ($e_{\mbox{\scriptsize{J}}}\sim0.05$) among the planets in the outer solar system, for the observed Jupiter Trojans with the number in excess of 7000, we statistically analyzed the distribution of their orbital orientations. 

It is interesting to find that, the L4 and L5 Trojan swarms have longitudes of perihelia $\varpi$  gathered around $\varpi_{\mbox{\scriptsize{J}}}+60^{\circ}$ and $\varpi_{\mbox{\scriptsize{J}}}-60^{\circ}$ ($\varpi_{\mbox{\scriptsize{J}}}$ is the longitude of perihelion of Jupiter), respectively. And for the $\varpi$ bins with the width of $30^{\circ}$, either of these two swarms has a number fraction peaked over 0.2, which is much larger than the fraction of $\sim0.083$ resulted from the uniform distribution, at the $5\sigma$ confidence level from Poisson statistics. This suggests that the apsidal asymmetric-alignment of Jupiter Trojans is noticeable. By integrating the orbits of Jupiter Trojans up to 1 Myr, the clustering in $\varpi$ can persist over time, thus this orbital characteristic should be robust but not due to the observational bias. The apsidal clustering of the Trojans is supposed to be caused by Jupiter which possesses a relatively large eccentricity of $e_{\mbox{\scriptsize{J}}}\sim0.05$. However, the distribution of the longitudes of ascending nodes for the observed Trojans is rather uniform, since the inclination of Jupiter is very low.

Next, in the framework of the spatial elliptical restricted three-body problem, we developed a theoretical approach to understand the clustering in $\varpi$ for the Jupiter Trojans. The averaged Hamiltonian with three degrees of freedom is derived to describe the dynamics of the 1:1 mean motion resonance. By taking the resonant angle to be $\psi = 60^{\circ}$ ($-60^{\circ}$) for the L4 (L5) Trojan swarm, the Hamiltonian can be reduced to a two degrees of freedom system, in which the momenta are in terms of the Trojan's eccentricity $e$ and inclination $i$. Then we provide a global view of the dynamical aspects for the co-orbital motion in the phase space ($e$, $\Delta\varpi$), where $\Delta\varpi=\varpi-\varpi_{\mbox{\scriptsize{J}}}$. The portraits show that there are equilibrium points at ($e\sim e_{\mbox{\scriptsize{J}}}, \Delta\varpi\sim\pm60^{\circ}$), either for the coplanar or inclined cases with $i\le40^{\circ}$. This can nicely account for the distinctive apsidal asymmetric-alignment of Jupiter Trojans as revealed above. We further note that the apsidally aligned islands of libration exist only for the Trojan-like orbits with $e<e_c\sim0.1$, and this is essentially determined by Jupiter's eccentricity $e_{\mbox{\scriptsize{J}}}\sim0.05$. However,  only a small fraction of the known Jupiter Trojans satisfy this eccentricity condition.

\begin{figure}
 \hspace{0 cm}
  \includegraphics[width=8.5cm]{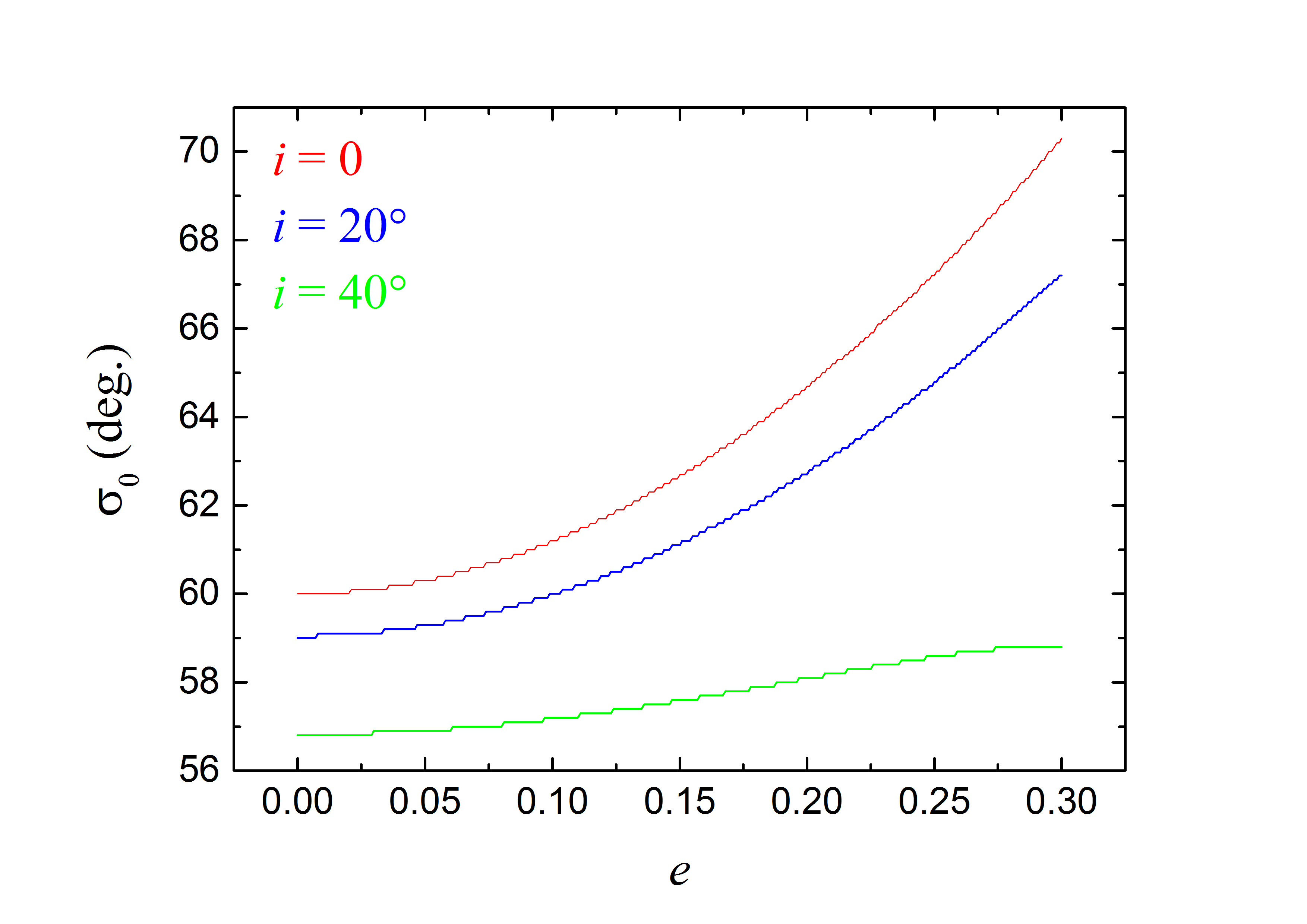}
  \caption{The libration center $\psi_0$ of the 1:1 mean motion resonance with respect to the eccentricity $e$ for the inclination $i=0$ (red), $20^{\circ}$ (blue) and $40^{\circ}$ (green),  determined from the circular restricted three-body model.}
  \label{LC}
\end{figure}

For better understanding the influence of the Trojan's eccentricity, the proper value $e_p$ and the maximum value $e_{\mbox{\scriptsize{Max}}}$ are computed in their 1 Myr evolution. Previous studies show that the two Lagrangian points have almost the identical dynamical features concerning the orbits around them, thus we then consider only the L4 Trojans. We introduced a time index $T_{frac}$ to measure the fraction of time that the L4 swarm has $\varpi$ placed in the range of $|\Delta\varpi-60^{\circ}|<\delta=60^{\circ}$. The results show that, for the population with $e_p<0.1$, the derived time fraction $T_{frac}$ is much larger than the value corresponding to a uniform variation of $\varpi$ in time. It suggests that although some objects evolve to the orbits with osculating eccentricities exceeding 0.1 (i.e., having $e_{\mbox{\scriptsize{Max}}}>0.1$), and accordingly could temporarily escape the libration islands of $\Delta\varpi$, they actually spend most of their time in the said $\Delta\varpi$ range. Because the more eccentric Trojans with $e_p>0.1$ are very few in number, the entire Trojan population would display the obvious apsidal alignment.

Finally, we proceed to investigate whether Jupiter Trojans could reach this peculiar orbital structure in the long-term evolution. Initially, tens of thousands of test Trojans around the L4 point are uniformly distributed between $\varpi=0-360^{\circ}$, given $e=0-0.3$ and $i=0-40^{\circ}$. Then by numerical simulations, we compute the evolution of these samples, sculpted by the eccentric Jupiter and the other three planets in the outer solar system. The orbital distributions obtained at different epochs show that, as time passes, the clustering of the simulated Trojans around $\Delta\varpi=60^{\circ}$ becomes more and more prominent. And at the end of the $10^9$ yr integration, the number fraction of the apsidally aligned objects is peaked about 0.2, for the $\Delta\varpi$ bins with the width of $30^{\circ}$, similar to what is observed in the real Trojan population. A series of additional runs, for which the test Trojans have initial $\varpi$ obeying Gaussian distributions with different means, have also been carried out.  We find that the similar apsidal alignment around $\Delta\varpi=60^{\circ}$ can always be reproduced.

In this work, our theoretical approach built on the restricted three-body model suffers from a minor problem: for the 1:1 resonant angle $\psi$ of the co-orbital motion, the libration center $\psi_0$ does not fix at the classical Lagrangian point of $60^{\circ}$ (L4) or $-60^{\circ}$ (L5), but it actually varies with the Trojans' eccentricities $e$ and inclinations $i$. By expanding the disturbing function to the fourth order in $e$ and $i$, \citet{namo00} studied the orbits librating around L4 or L5 with small amplitudes, and they showed that the displacements of equilibrium points from the equilateral configuration increase as a function of increasing $e$. The displaced equilibrium points were also calculated for the coplanar configuration of two massive planets in the co-orbital motion \citep{giup10}. As shown in Fig. 7 of their work, given a ratio $m_2/m_1\ge1$ of planetary masses, the equilibrium value of $\psi$ (i.e., $\psi_0$) for the L4 solutions increases with the eccentricity $e_1$ of the less massive planet. Considering the limit of the mass ratio $m_2/m_1\rightarrow\infty$, i.e., in the restricted three-body problem, this trend is the same as that discovered by \citet{namo00}. By using the semi-analytical method developed in our previous works for the resonant Kuiper belt objects \citep{li14a, li14b, li2020}, the libration center $\psi_0$ for the L4 co-orbital motion is determined as a function of the massless Trojan's $e$ for different $i$. Considering the observed range of $e\sim0-0.3$ and $i\sim0-40^{\circ}$, Fig. \ref{LC} shows that $\psi_0$ could change between $57^{\circ}$ and $70^{\circ}$, indicating that the equilibrium point within the apsidal alignment islands (see Fig. \ref{theory1}a) may shift no further than $10^{\circ}$ from $\Delta\varpi=60^{\circ}$. Since the L4 Trojans are mainly clustered in a much wider range of $\Delta\varpi=0-120^{\circ}$, the $\lesssim10^{\circ}$ displacement of $\Delta\varpi$ would not affect our conclusions. This argument can be supported by the apsidal alignment of Jupiter Trojans from both the observation and numerical reproduction.

%______________________________________________________________

\section*{Acknowledgments}

This work was supported by the National Natural Science Foundation of China (Nos. 11973027, 11933001, 12073011 and 11603011), and National Key R\&D Program of China (2019YFA0706601). We would also like to express our sincere thanks to the anonymous referee for the valuable comments.

\section*{Data Availability}

The data underlying this article are available in the article and in its online supplementary material.

\end{document}